\documentclass[fleqn,usenatbib]{mnras}
\pdfoutput=1
\usepackage[T1]{fontenc}
\usepackage{ae,aecompl}

\usepackage{graphicx}
\usepackage{amsmath}
\usepackage{amssymb}
\usepackage[ddmmyyyy]{datetime}
\usepackage{blindtext}
\usepackage{xcolor}

\graphicspath{{figures/}}

\hypersetup{breaklinks, psdextra, pdfencoding=auto}

\DeclareMathOperator{\sinc}{sinc}

\title[21cm-LAE Cross-Power Spectrum]{Probing delayed-end reionization histories with the 21cm-LAE cross-power spectrum}

\author[L. Weinberger et al.]{
Lewis H. Weinberger$^{1}$\thanks{Email: lewis.weinberger@ast.cam.ac.uk}
Girish Kulkarni$^{2}$ and Martin G. Haehnelt$^{1}$
\\
$^{1}$ Institute of Astronomy and Kavli Institute for Cosmology, University
of Cambridge, Madingley Road, Cambridge CB3 0HA, UK \\
$^{2}$ Department of Theoretical Physics, Tata Institute of Fundamental Research,
Homi Bhabha Road, Mumbai 400005, India
}

\date{Accepted XXX\@. Received YYY; in original form ZZZ}
\pubyear{2020}

\begin{document}
\label{firstpage}
\pagerange{\pageref{firstpage}--\pageref{lastpage}}
\maketitle

\begin{abstract}
  We model the 21-cm signal and LAE population evolution during the epoch of reionization in order to predict the 21cm-LAE cross-power spectrum. We employ high-dynamic-range simulations of the IGM to create models that are consistent with constraints from the CMB, Lyman-$\alpha$ forest and LAE population statistics. Using these models we consider the evolution of the cross-power spectrum for a selection of realistic reionization histories and predict the sensitivity of current and upcoming surveys to measuring this signal. We find that the imprint of a delayed-end to reionization can be observed by future surveys, and that strong constraints can be placed on the progression of reionization as late as $z=5.7$ using a Subaru-SKA survey. We make predictions for the signal-to-noise ratios achievable by combinations of Subaru/PFS with the MWA, LOFAR, HERA and SKA interferometers for an integration time of 1000 hours. We find that a Subaru-SKA survey could measure the cross-power spectrum for a late reionization at $z=6.6$ with a total signal-to-noise greater than 5, making it possible to constrain both the timing and bubble size at the end of reionization. Furthermore, we find that expanding the current Subaru/PFS survey area and depth by a factor of three would double the total signal-to-noise.
\end{abstract}




\section{Introduction}
During the cosmic period of history known as the epoch of reionization (EoR) the first stars and galaxies altered the large-scale ionization state of the Universe, with consequences persisting down to the present epoch \citep{2018PhR...780....1D}. In particular this period represents a phase transition in which the primordial neutral hydrogen in the intergalactic medium (IGM) --- left over from recombination \citep{2001PhR...349..125B} --- was ionized by sources such as galaxies and possibly early quasars \citep{2001ARA&A..39...19L}. Observations of the IGM in the high redshift Universe \citep[such as the Lyman-$\alpha$ forest seen in sightlines to $z=5$--6 quasars,][]{2015MNRAS.447..499M} constrain the neutral hydrogen fraction to be $\overline{x}_{\rm HI} < 0.1$ by $z\sim 6$. The reionization process involved radiative transfer across a range of scales, from within galaxies \citep{2019MNRAS.483.1029K} to the farthest reaches of the IGM \citep{2000ApJ...530....1M}. Its impacts included photoheating of the IGM \citep{1994MNRAS.266..343M} and the inhibition of star formation in dwarf galaxies \citep{1992MNRAS.256P..43E,2019arXiv190511414K}. From a theoretical standpoint, reionization can indirectly give us constraints on the first populations of stars and early galaxies \citep{2015MNRAS.451.2030D}.

A number of methods have been established to measure reionization. One direct probe of the ionization state of the IGM is through the 21-cm signal from neutral hydrogen \citep{2012RPPh...75h6901P,2011MNRAS.411..955M,2006PhR...433..181F}. Radiation at a rest-frame wavelength of 21 cm is emitted and absorbed by neutral hydrogen via the hyperfine transition of the ground state \citep{1952ApJ...115..206W}. It should be possible to observe the 21-cm signal from the high redshift IGM, relative to the cosmic microwave background (CMB) radiation, using radio interferometry \citep{2000ApJ...528..597T}. Observations of the 21-cm signal are complicated by foreground radio signals \citep{2016MNRAS.462.3069S,2015aska.confE...5C} which must be removed or avoided \citep{PhysRevD.90.023019}.

Another indirect probe of reionization is the population statistics of Lyman-$\alpha$ emitters (LAEs), such as the luminosity and angular clustering functions. Intervening neutral hydrogen gas can obscure the Lyman-$\alpha$ emission from distant LAEs, so we expect to see an attenuation of the luminosity function at redshifts within the EoR \citep{1965ApJ...142.1633G,2000ApJ...542L..69M,2007MNRAS.377.1175D}. Interpreting LAE observations in the context of reionization is challenging because the radiative transfer processes occuring within the galaxy's ISM --- including the interaction with dust --- is degenerate with attenuation by the IGM \citep{2014MNRAS.441.2861H,2015MNRAS.450.4025H}. Alongside the attenuating effect on the luminosity function, another imprint of reionization on the high-$z$ LAE population is the enhancement of angular clustering. We expect LAEs to reside within the overdensities that sit inside the first ionized bubbles, and this spatial correlation can enhance the observed clustering signal \citep{2007MNRAS.381...75M}.

There has been extensive theoretical work modelling both the 21-cm signal \citep[including][]{2018ApJ...867...26B,2017MNRAS.468.3869S,2017arXiv170104408K,2016MNRAS.457.1550H,2016MNRAS.463.2583K,2011MNRAS.411..955M,2006ApJ...653..815M} and the LAE population \citep[including][]{2018arXiv180607392L,2018arXiv180100067I,2019MNRAS.485.1350W,2018arXiv180101891M,2016MNRAS.463.4019K,2015MNRAS.450.4025H,2015ApJ...812..123G,2014MNRAS.441.2861H,2014ApJ...794..116Z,2013MNRAS.428.1366J,2007MNRAS.377.1175D} during the epoch of reionization. The epoch of reionization has been explored with analytic prescriptions \citep[such as][]{2009CSci...97..841C,2004ApJ...613....1F}, using semi-numerical modelling \citep[such as][]{2011MNRAS.411..955M,2014MNRAS.443.2843M}, and full numerical simulation \citep[such as][]{2007ApJ...671....1T,2018MNRAS.479..994R,2015MNRAS.453.2943C}. The challenge of modelling reionization is to try to capture the physical processes which occur on extremely different scales: from the large-scale ionized bubble structure \citep{2007MNRAS.377.1043M,2007ApJ...654...12Z,2018MNRAS.473.2949G}, to the small-scale self-shielding of neutral gas clumps \citep{2013MNRAS.430.2427R,2017arXiv170706993C}. Accurately modelling this large dynamic range is particularly important for the 21-cm signal \citep{2016MNRAS.463.2583K}. Similarly predicting the effect of reionization on the observatibility of LAEs is complicated by resonant radiative transfer of the Lyman-$\alpha$ emission within the galaxy and the wider halo environment \citep{2017ApJ...839...44S,2014PASA...31...40D}. In this work we employ the 21-cm model used in \citet{2016MNRAS.463.2583K}, combined with the reionization histories and LAE modelling described in \citet{2019MNRAS.485.1350W}.

\begin{figure}
  \includegraphics[width=0.5\textwidth]{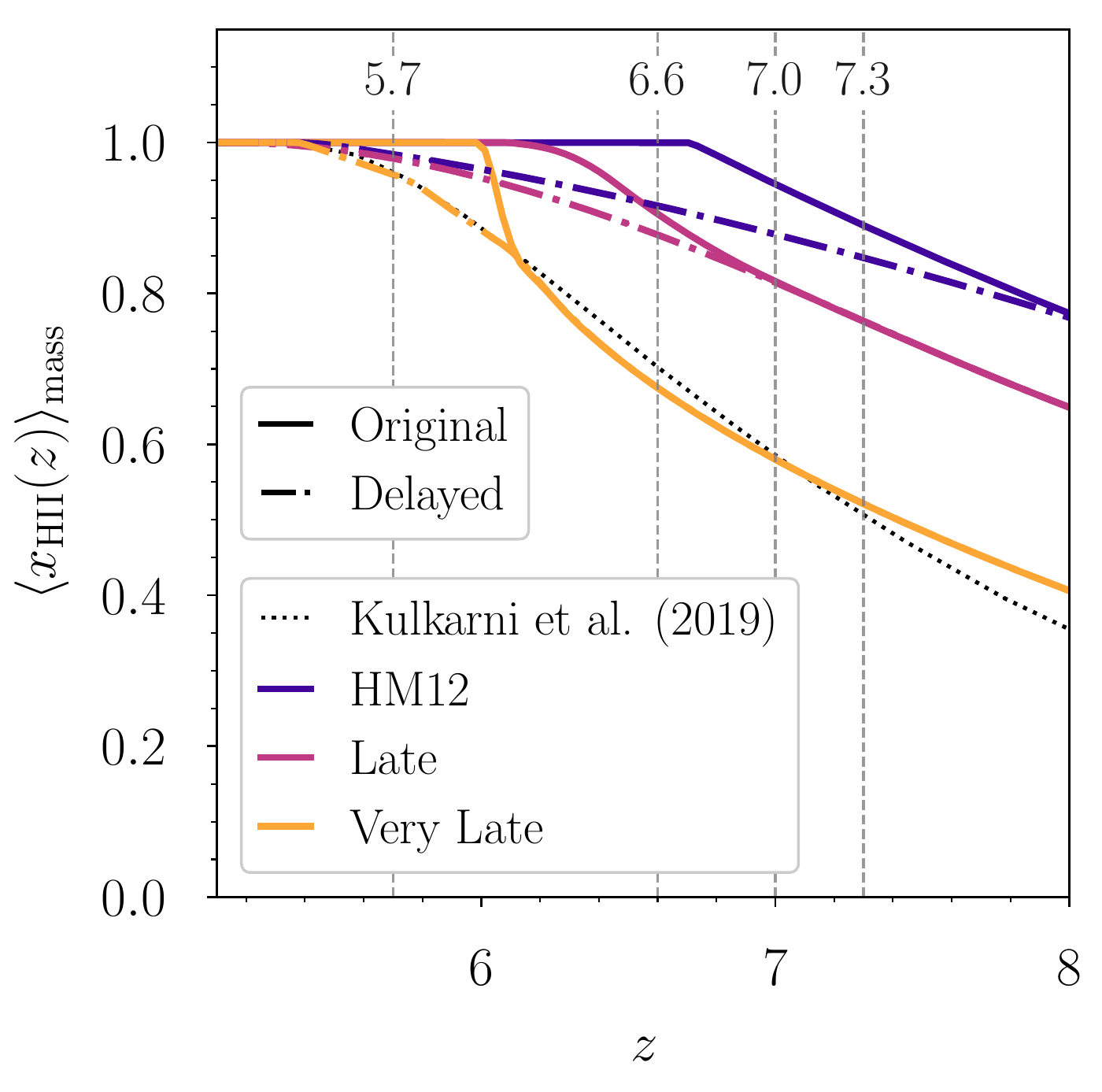}
  \caption{The mass-averaged ionized fraction for the six reionization histories considered in this work, showing the \emph{original} models with solid lines and the \emph{delayed-end} models with dot-dashed lines. The positions of the Subaru narrow-band filter centres are shown with the grey dashed lines. For comparison the reionization history considered in \citet{2018arXiv180906374K} is shown with the black dotted line.}
  \label{fig:1}
\end{figure}

The cross-correlation of the 21-cm signal and the LAE population at high redshifts has been considered previously by a number of authors. Early work considered the possibility of measuring the cross-power spectrum \citep{2007ApJ...660.1030F,2009ApJ...690..252L}, predicting observational sensitivities for instruments such as the Murchison Widefield Array \citep[MWA,][]{2014MNRAS.438.2474P}, the LOW Freqency ARray \citep[LOFAR,][]{2013MNRAS.432.2615W,2016MNRAS.457..666V} and the Square Kilometer Array \citep[SKA,][]{2016MNRAS.459.2741S}. The cross-correlation of these two observables is not subject to the radio foregrounds that plague the interpretation of individual 21-cm observations. Many theoretical approaches use computational simulations of the IGM to model the cross-correlation signal, but it has also been considered analytically \citep[for example, in][]{2017arXiv170107005F}. Recent work has explored the synergy of the SKA and Subaru, considering both the power spectrum and the correlation function. These studies found that SILVERRUSH type Subaru surveys combined with the SKA are optimal for distinguishing an IGM with average neutral fractions of 50\%, 25\% and 10\% at $z=6.6$ \citep{2018MNRAS.479L.129H,2017ApJ...836..176H}. These works also established that deeper and wider LAE surveys can improve the sensitivity to measuring the cross-power spectrum \citep{2019arXiv191002361K,2018MNRAS.479.2767Y,2018MNRAS.479.2754K}. Another interesting area of research is the cross-correlation of the 21-cm signal with other types of galaxies, for example [\ion{O}{III}] emitters \citep{2019arXiv190610863M}, or with intensity maps of other emission lines \citep{2017ApJ...849...50N}.

In this work we employ our empirical models with the high-dynamic-range Sherwood simulations to test the effect of six reionization histories on the cross-correlation evolution, including the delayed-end history that has successfully explained the large opacity fluctuations in the Ly$\alpha$ forest \citep{2018arXiv180906374K,2019arXiv190512640K,2019arXiv191003570N}. We make predictions for a selection of observational surveys using updated parameters, but following the framework established by earlier work \citep[][in particular]{2007ApJ...660.1030F,2009ApJ...690..252L}. Our aim is to determine whether observations of 21cm-LAE cross correlations can provide additional evidence to support a delayed end to reionization. In order to determine this, we explore how sensitive the different surveys are to measuring the 21cm-LAE cross power spectrum in a delayed reionization scenario, and also to what extent this scenario can be distinguished from earlier reionization histories.

This paper is structured as follows: in section~\ref{sec:method} we detail the setup of our simulations and theoretical models; in section~\ref{sec:results} we present our results for the cross-power spectrum and our predictions for observational sensitivities; in section~\ref{sec:discussion} we discuss the implications of these results; and finally we conclude in section~\ref{sec:conclusions}. In particular, section~\ref{sec:method} is split into subsections describing: our simulations (\S\ref{sec:reionization}), our model for the 21-cm signal (\S\ref{sec:21cm}), our LAE population model (\S\ref{sec:LAE}), and the definitions we use for the power spectra calculations (\S\ref{sec:ps_details}). Alongside this we include a more detailed description of the numerical implementation of our power spectrum calculation in Appendix~\ref{appendix:ps_details}, as well as derivations of our sensitivity modelling in Appendix~\ref{appendix:surveys}.

\section{Method}
\label{sec:method}
\subsection{Simulating reionization}
\label{sec:reionization}
In this sub-section we summarise the key features of our simulation set-up. We note that it is the same as that used in \citet{2019MNRAS.485.1350W,2018MNRAS.tmp.1485W}, and we refer the reader to those papers for further detail.

\subsubsection{Cosmological simulations}
The basis for our theoretical modelling is the Sherwood suite of cosmological hydrodynamic simulations \citep{2017MNRAS.464..897B}, designed to achieve a high-dynamic-range and effectively capture the behaviour of the low-density IGM.\@ In this work we make use of a periodic box of side $L = 320$ cMpc/h, run using the \textsc{P-Gadget-3} \citep{GADGET-2,2001NewA....6...79S} SPH code with $N = 2\times 2048^3$ particles ($M_{\rm DM} = 2.75\times 10^8$ h$^{-1} M_\odot$). A Friends-of-Friends algorithm was used to identify dark matter haloes on-the-fly. This simulation used the \citet{2014A&A...571A..16P} cosmological paramaters: $h=0.678$, $\Omega_m = 0.308$, $\Omega_\Lambda = 0.692$, $\Omega_b = 0.0482$, $\sigma_8 = 0.829$, $n = 0.961$, and $Y_{\mathrm{He}} = 0.24$. When calculating the 21-cm brightness temperature, the SPH kernel was used to interpolate particles onto a uniform grid.

The Sherwood suite has been used to model a variety of phenomena such as the 21-cm reionization signal \citep{2016MNRAS.463.2583K,2017arXiv170104408K}, the LAE evolution at high redshift \citep{2018MNRAS.tmp.1485W,2019MNRAS.485.1350W}, the detectability of Ly-$\alpha$ emission in the cosmic web \citep{2019arXiv190506954W}, a late reionization and the opacity fluctuations in the Ly-$\alpha$ forest \citep{2019arXiv190512640K,2018arXiv180906374K}, and the detectability of [\ion{C}{II}] line intensity mapping \citep{2019MNRAS.485.3486D}.

\subsubsection{Modelling the IGM ionization structure}
In order to explore the effect of reionization on the 21cm-LAE cross-power spectrum we implement the calibrated reionization modelling of \citet{2015MNRAS.452..261C}. This involves first generating the large-scale ionization structure using an excursion set approach \citep{2004ApJ...613....1F,Mesinger:2007pd,2009MNRAS.394..960C,
2011MNRAS.411..955M, 2010MNRAS.406.2421S, 2016MNRAS.457.1550H}, followed by a calculation of the ionization state of the small-scale ionized regions which are assumed to be in photoionization equilibrium. For a given reionization history which maps the average ionized fraction of the IGM, $\bar{x}_{\rm HII}(z)$, the two steps of large- and small-scale ionization structure are calibrated to find a consistent UV-background photoionization rate, $\Gamma_{\rm HI}(z)$. This allows us to impose arbitrary reionization histories specified by $\bar{x}_{\rm HII}(z)$, and calculate the resulting ionization structure self-consistently across the scales probed by the simulation.

As part of the photoionization equilibrium calculation, we employ the self-shielding prescription of \citet{2017arXiv170706993C}. This is a redshift-dependent modification of the \citet{2013MNRAS.430.2427R} parametrization, which accounts for changes to the local photoionization rate in the proximity of overdense self-shielded gas.

\begin{figure}
  \includegraphics[width=0.5\textwidth]{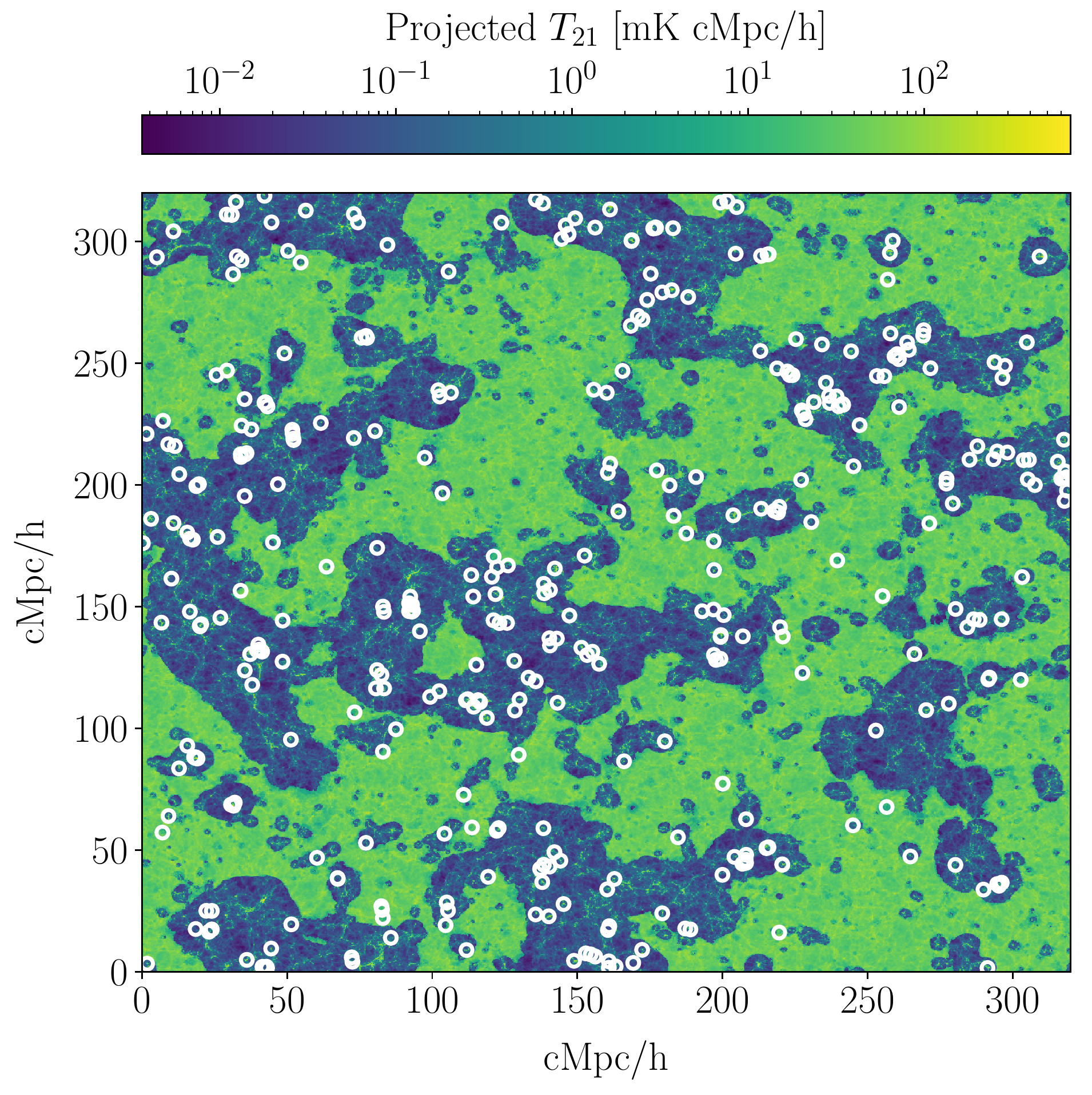}
  \caption{The 21-cm brightness temperature projected over a thin 2.5 cMpc/h slice of the simulation at $z=6.604$ for the delayed-end Very Late model. Overplotted with white empty circles are the positions of simulated LAEs within this thin slice.}
  \label{fig:2}
\end{figure}

\subsubsection{Reionization histories}
We test six physically motivated reionization histories in this work, using the same models as in \citet{2019MNRAS.485.1350W}. These histories are based on the original three reionization histories explored by \citet{2015MNRAS.452..261C}:
\begin{itemize}
\item {\bf HM12}, based on the minimal reionization model of \citet{2012ApJ...746..125H}. This is the earliest reionization model considered in this work, with reionization ending by $z=6.7$.
\item {\bf Late}, similar to the HM12 evolution but shifted in redshift so that it ends at a later time of $z=6$.
\item {\bf Very Late}, a further variation of the Late evolution, in which the end of reionization remains at $z=6$ but the gradient ${\rm d}\overline{x}_{\rm HII}/{\rm d}z$ is altered to be more abrupt.
\end{itemize}
We use these three models (hereafter referred to as the \emph{original} models), as well as three modified versions (referred to as the \emph{delayed-end} models) in which the end of reionization is delayed to $z\sim 5.3$.

In particular we note that the delayed-end Very Late model has a very similar evolution to the reionization histories considered in \citet{2018arXiv180906374K} \& \citet{2019arXiv190512640K}, which successfully simulated the opacity fluctuations seen in observations of the Ly-$\alpha$ forest \citep{2015MNRAS.447.3402B}. In \citet{2019MNRAS.485.1350W}, the LAE evolution predicted by this delayed-end Very Late model was found to be consistent with observations of the equivalent width distribution, luminosity function and angular correlation function as measured by the SILVERRUSH survey \citep{2017arXiv170501222K,2018PASJ...70S..14S,2018arXiv180505944I}. We also note that the lower value of $\tau = 0.054 \pm 0.007$ measured by \citet{2018arXiv180706209P} is consistent with a late end to reionization. This reionization model therefore offers a neutral fraction evolution that is consistent with many independent observational constraints on reionization.

\begin{table}
\caption{Observational selection thresholds \citep[as in][]{2019MNRAS.485.1350W} used to generate mock observed samples.}
\label{tab:thresholds}
\begin{tabular}{l|c|c|c}
\hline
Based on survey & $z$ & REW$_\mathrm{min}$ [\AA] & $L_\mathrm{Ly\alpha,min}$ [erg/s] \\ \hline
 \citet{2017arXiv170501222K} & 5.7 & 10 & 6.3$\times10^{42}$ \\ \hline
 \citet{2017arXiv170501222K} & 6.6 & 14 & 7.9$\times10^{42}$ \\ \hline
 \citet{2017arXiv170302501O} & 7.0 & 10 & 2$\times10^{42}$ \\
 \citet{2018arXiv180505944I} &     &    &                  \\ \hline
 \citet{2014ApJ...797...16K} & 7.3 & 0  & 2.4$\times10^{42}$ \\ \hline
\end{tabular}
\end{table}

The average ionized fraction evolution for each of our six reionization histories can be seen in Figure~\ref{fig:1}.

\begin{table*}
    \caption{Survey parameters for modelling observational sensitivities, as detailed in Appendix~\ref{appendix:surveys}. For all the 21-cm surveys we assume an integration time of $t=1000$ hours, a bandpass of $B=8$ MHz and a spectral resolution of $\Delta\nu=50$ kHz, hence the available parallel $k$-modes at $z=6.6$ lie in the interval between $k_{\rm \parallel, min} = 0.056$ cMpc$^{-1}$ and $k_{\rm \parallel, max} = 8.983$ cMpc$^{-1}$.}
    \label{tab:surveys}
    \begin{tabular}{l|c|c|c|c|c}
      \hline
      Parameters & MWA$^a$ & LOFAR$^b$ & HERA$^c$ & SKA1-low$^d$ & Subaru/PFS$^e$ \\
      \hline
       Effective collecting area, $A_e^\dagger$ [m$^2$] & 21.5  & 512.0 & 77.6 & 497.4  & -- \\
       Number of antennae, $N_a$ & 256$^\ddagger$ & 24 & 350 & 512 & -- \\
       Core radius, $r_{\rm core}$ [m] & 100 & 160 & 150 & 350 & -- \\
       Maximum radius, $r_{\rm max}$ [km] & 3.5 & 2.0 & 0.45 & 6.4 & -- \\
       Minimum baseline, $b_{\rm min}$ [m] & 7.7 & 68.0 & 14.6 & 35.0 & -- \\
       Maximum baseline, $b_{\rm max}$ [km] & 5.3 & 3.5 & 0.876 & 12.8 & -- \\
       Minimum transverse $k$-mode at $z=6.6$, $k_{\rm \bot, min}$ [cMpc$^{-1}$] & 0.003 & 0.031 & 0.007 & 0.016 & -- \\
       Maximum transverse $k$-mode at $z=6.6$, $k_{\rm \bot, max}$ [cMpc$^{-1}$] & 2.405 & 1.588 & 0.397 & 31.761 & -- \\
      \hline
       Spectroscopic redshift error, $\sigma_{z}$ & -- & -- & -- & -- & 0.0007 \\
      \hline
       Survey volume at $z=6.6$, $V_{\rm survey}$ [cMpc$^3$] & $1.6 \times 10^9$ & $6.6 \times 10^7$ & $4.3 \times 10^8$ & $6.8 \times 10^7$ & $2.3\times 10^7$\\
       Field-of-view, $\Omega^{\dagger}$ [deg$^2$] & 610 & 26 & 169 & 26 & 27 \\
      \hline
    \end{tabular}
  \vspace{0.1cm}

  {\raggedright \footnotesize
    $^a$ Extended configuration, based on \citet{2018PASA...35...33W,2013PASA...30....7T}.

    $^b$ NL-Core configuration, based on \citet{2013A&A...556A...2V}.

    $^c$ Hera-350 configuration, based on \citet{2017PASP..129d5001D}.

    $^d$ V4A core configuration, based on \citet{SKA-SCI-LOW-001, 2015aska.confE...4C}.

    $^e$ We assume a narrowband ($\Delta z = 0.1$) survey geometry, based on \citet{2018PASJ...70S...1M,2014PASJ...66R...1T}.

    $^\dagger$ Observing at 150 MHz. We assume a frequency dependence of $A_{e}(\nu)/A_{e}(\nu=150\rm MHz) = (150 / \nu)^2$.

    $^\ddagger$ The Phase II upgrade gives the MWA a total of 256 tiles, however with the existing receivers and correlator it is only possible to use 128 at any one time \citep{2018PASA...35...33W}.

  }
\end{table*}

\subsection{Simulating 21-cm emission}
\label{sec:21cm}
We follow the approach of \citet{2016MNRAS.463.2583K} to model the 21-cm brightness temperature using the Sherwood simulations. We assume that the spin temperature of hydrogen is much greater than the cosmic microwave background (CMB) temperature, which should be valid for the redshifts considered in this work where the average ionized fraction of the IGM is greater than a few percent \citep{2012RPPh...75h6901P}. We can then calculate the brightness temperature of the 21-cm line relative to the CMB as a function of the \ion{H}{I} fraction, $x_{\rm HI}$, and density contrast, $\Delta = \rho/\bar{\rho}$, using \citep{2006PhR...433..181F},
\begin{equation}
  T_{21} (\textbf{x}) = \overline{T}_{21} \: x_{\rm HI}(\textbf{x}) \: \Delta(\textbf{x}),
\end{equation}
where the brightness temperature of the gas at mean density is \citep{2009MNRAS.394..960C},
\begin{equation}
\overline{T}_{21} \simeq 22\:{\rm mK}\: \left[(1 + z)/7\right].
\end{equation}
In order to calculate this for our simulations we use the SPH kernel to interpolate the SPH particles onto a uniform grid with 1024 cells on a side (cell size $L_{\rm cell} = 312.5$ ckpc/h) allowing us to calculate the density contrast at a given cell position, $\textbf{x}$. We apply the reionization history modelling as described in section~\ref{sec:reionization} to determine the \ion{H}{I} fraction in each grid cell.

\subsection{Simulating Lyman-\texorpdfstring{$\alpha$}{alpha} emitters}
\label{sec:LAE}

We employ the LAE model developed in \citet{2019MNRAS.485.1350W}\footnote{A Python implementation of this model is publicly available at \url{https://github.com/lewis-weinberger/slap}.}. As in section~\ref{sec:reionization} we will now summarise the key features of this model, and refer the reader to the original work for further details. An LAE population is generated within the simulation in the following stages:
\begin{itemize}
  \item \textbf{Identify LBG galaxies}: The first stage of the model is to determine which dark matter haloes host LBG-like galaxies. We abundance match the dark matter halo mass function with observed UV luminosity functions at $z=5.9$, using a duty-cycle following the prescription of \citet{2010ApJ...714L.202T}. This duty cycle is parameterised by $\Delta t$, which gives a measure of the time period over which the observability of the LBG population is evolving. This creates a mapping between halo mass, $M_h$, and LBG UV luminosity, $L_{\rm UV}$. We assume this mapping applies to other redshifts, or in other words that the LBG population evolution is driven by the underlying halo population evolution (at high redshifts). This means we can apply the mapping at other redshifts to derive a population of LBGs (a subset of the halo population) with assigned UV luminosity, $L_{\rm UV}$.
  \item \textbf{Identify LAE galaxies}: The second stage of the model identifies which of the LBG population has sufficient Lyman-$\alpha$ emission to be observed as an LAE\@. We employ the Lyman-$\alpha$ rest-frame equivalent width (REW) distribution model of \citet{2012MNRAS.419.3181D}, which assigns a conditional probability for an LBG having a certain Lyman-$\alpha$ REW given its UV magnitude, $P({\rm REW}|M_{\rm UV})$. In particular we assume this model describes the \emph{intrinsic} REW distribution of the LBG population, when attenuation of Lyman-$\alpha$ radiation by the IGM is neglected. For each LBG we sample from this distribution to derive its intrinsic REW (and hence also its intrinsic Lyman-$\alpha$ luminosity), thus generating an LAE population.
  \item \textbf{Apply observational selection}: The final stage of the model is to determine which LAEs would be observed, including the attenuation of their emission by the IGM\@, given the selection function of a particular survey. We employ observational limits on the LAE REW and $L_{\rm Ly\alpha}$ based on the SILVERRUSH surveys, shown in Table~\ref{tab:thresholds}. To calculate the radiative transfer of Lyman-$\alpha$ emission from the LAEs to the observer, we extract sightlines through the simulation (with ionization structure derived as in section~\ref{sec:reionization}). We assume the complex radiative transfer within the halo produces a double-peaked emission profile, of which only the red peak will escape through the IGM\@. We model this intrinsic emission profile as a Gaussian peak \citep{2015MNRAS.452..261C}, with a velocity offset that scales with the host halo virial velocity $\Delta v \propto v_{\rm circ}$. For the radiative transfer throughout the rest of the IGM, we calculate the effect of scattering out of the line-of-sight using the $e^{-\tau}$ approximation \citep{2011ApJ...728...52L}. We then calculate the Lyman-$\alpha$ transmission fraction, $T^{\rm IGM}_{\rm Ly\alpha}$, along a single sightline to each LAE so that we can finally determine their observed luminosity, $L^{\rm obs}_{\rm Ly\alpha} = T^{\rm IGM}_{\rm Ly\alpha} L^{\rm int}_{\rm Ly\alpha}$. Note we are therefore neglecting any intrinsic or dust-driven evolution of the emission profile.
\end{itemize}
We apply our LAE model to four snapshots of our Sherwood simulation, at redshifts $z=5.756, 6.604, 6.860, 7.444$  which are close to the central redshifts measured by the NB816, NB921, NB973, NB101 filters on the HSC of the Subaru telescope. This generates mock LAE populations which could be observed with narrowband observations using the Subaru telescope.

Importantly the key free parameters in the LAE model are the duty cycle parameter, $\Delta t$, and the velocity offset, $\Delta v$. We calibrate these parameters at $z=5.7$, such that our mock population matches the observed REW distribution, luminosity function and angular correlation function as measured by \citet{2018PASJ...70S..14S,2017arXiv170501222K,2017arXiv170407455O}. These calibrated parameters are assumed to be fixed for the other redshifts, such that any evolution in the LAE population is due to the underlying halo population and IGM neutral fraction evolution. The velocity offset, $\Delta v = a\: v_{\rm circ}$, is calibrated to,
  \begin{equation}
    a =
    \begin{cases}
        1.5 & \mathrm{Original}, \\
        1.8 & \mathrm{Delayed},
    \end{cases}
  \end{equation}
for the two different reionization history cases.
In \citet{2019MNRAS.485.1350W} it was found that --- after calibrating at $z=5.7$ --- these models compare favourably to the observed LAE populations in the SILVERRUSH survey, in particular predicting the evolution of the luminosity function, equivalent width distribution and angular correlation function seen at the other narrowband redshifts.

In Figure~\ref{fig:2} we plot a slice of our simulation volume, showing a projection of the 21-cm brightness temperature and the positions of LAEs. We note that the LAEs are predominately found in the overdense regions which ionize first, and hence we mostly see LAEs in regions where the 21-cm signal is faint. Conversely the underdense regions which remain neutral for longer do not host many LAEs, and so we see a brighter 21-cm signal where there are fewer LAEs.

\begin{figure*}
  \includegraphics[width=\textwidth]{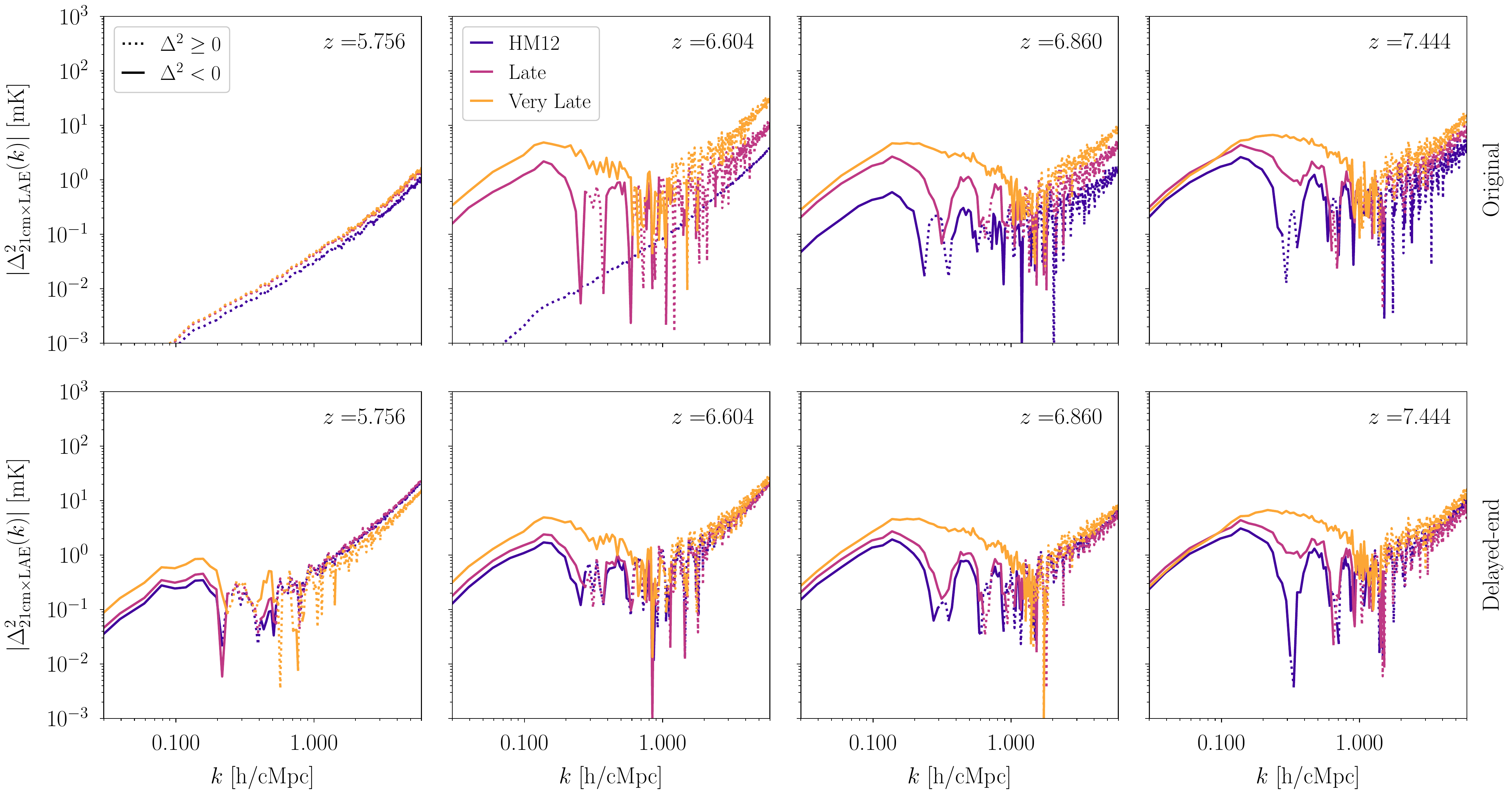}
  \caption{Cross power spectra for the reionization histories: HM12 (blue), Late (purple), Very Late (orange). The top panels show the original reionization histories, whilst the bottom panels show the delayed-end versions. From left to right the panels show increasing redshift, starting at $z=5.756$ on the left and ending with $z=7.444$ on the right. The dotted lines indicate where the power spectrum is positive ($\Delta^2 \geq 0$), whilst the solid lines indicate that the power spectrum is negative ($\Delta^2 < 0$).}
  \label{fig:3}
\end{figure*}

\subsection{Calculating the power spectrum}
\label{sec:ps_details}
We work with the brightness temperature $T_{\rm 21cm}(\mathbf{x})$ directly, and the LAE number density contrast defined as,
\begin{equation}
\delta_{\rm LAE} \equiv \frac{n_{\rm LAE}(\mathbf{x}) - \langle n_{\rm LAE} \rangle}{\langle n_{\rm LAE} \rangle},
\end{equation}
where $\langle.\rangle$ indicates the spatial average. As described above, the 21-cm brightness temperature, $T_{\rm 21cm}$, is modelled using our simulations on a uniform grid with $1024^3$ cells (with cell size 0.3125 cMpc/h). We interpolate the LAE population onto the same grid geometry using the Cloud-in-Cell (CIC) scheme. We can then calculate the cross-power spectrum as,
\begin{equation}
\label{eq:cross_ps}
\langle \tilde{T}_{\rm 21cm} (\mathbf{k_1})\tilde{\delta}_{\rm LAE} (\mathbf{k_2}) \rangle = (2 \pi)^3 \delta_{D}(\mathbf{k_1} + \mathbf{k_2}) P_{\rm 21cm \times LAE} (k),
\end{equation}%
where the tilde quantity $\tilde{X}(\mathbf{k})$ denotes the Fourier transform of the field $X(\mathbf{x})$, $\delta_{D}$ is the Dirac delta function and $P(k)$ is the power spectrum which depends only on $k=|\mathbf{k}|$ as a result of the statistical isotropy of the fields. Similarly the auto-power spectra are calculated using,
\begin{equation}
\label{eq:21_ps}
\langle \tilde{T}_{\rm 21cm} (\mathbf{k_1})\tilde{T}_{\rm 21cm} (\mathbf{k_2}) \rangle = (2 \pi)^3 \delta_{D}(\mathbf{k_1} + \mathbf{k_2}) P_{\rm 21cm} (k),
\end{equation}%
\begin{equation}
\label{eq:lae_ps}
\langle \tilde{\delta}_{\rm LAE} (\mathbf{k_1})\tilde{\delta}_{\rm LAE} (\mathbf{k_2}) \rangle = (2 \pi)^3 \delta_{D}(\mathbf{k_1} + \mathbf{k_2}) P_{\rm LAE} (k).
\end{equation}%
Further details of the numerical implementation of this calculation can be found in Appendix~\ref{appendix:ps_details}.

When presenting our results we will hereafter work with dimensionless power
spectra defined as,
\begin{equation}
\Delta^2 (k) \equiv \frac{k^3}{2 \pi^2} P(k).
\end{equation}
This quantity carries no length dimensionality, however given the conventions
defined in Eqs.~(\ref{eq:cross_ps})~\&~(\ref{eq:21_ps}) we note that $\Delta^2_{21}(k)$
has units of temperature squared, (mK)$^2$, whilst $\Delta^2_{\rm 21 \times LAE}(k)$
has units of temperature, mK.
We can also calculate the cross-correlation coefficient,
\begin{equation}
\label{eq:coeff}
r_{\rm 21cm \times LAE}(k) = \frac{P_{\rm 21cm \times LAE}(k)}{\sqrt{P_{\rm 21cm}(k)P_{\rm LAE}(k)}}.
\end{equation}

\subsection{Calculating observational sensitivities}
\label{sec:method_obs}
We follow the approach of \citet{2007ApJ...660.1030F} \& \citet{2009ApJ...690..252L} for exploring
the errors introduced in the measurements of the 21-cm signal and the LAE survey. For a particular \textbf{k}-mode measured along a given line-of-sight labelled by $\mu$ (the cosine of the angle between the line-of-sight and $\mathbf{k}$), and restricting ourselves to the upper-half \textbf{k}-plane, the variance on the cross-power spectrum is given by \citep{2019MNRAS.485.3486D},
\begin{multline}
{\rm var}[P_{\rm 21cm \times LAE}(k, \mu)] = \frac{1}{2}\left[P^2_{\rm 21cm \times LAE}(k, \mu) \right.\\ +
\left. \sqrt{{\rm var}[P_{\rm 21cm}(k, \mu)]\:{\rm var}[P_{\rm LAE}(k, \mu)]} \right],
\end{multline}
where the individual survey variances are given by,
\begin{equation}
\label{eq:21cmtot}
    {\rm var}[P_{\rm 21cm}(k,\mu)] = \left[P_{\rm 21cm}(k,\mu) + P^{\rm noise}_{\rm 21cm}(k,\mu) \right]^2,
\end{equation}
and,
\begin{equation}
\label{eq:laetot}
    {\rm var}[P_{\rm LAE}(k,\mu)] = \left[P_{\rm LAE}(k,\mu) + P^{\rm noise}_{\rm LAE}(k,\mu) \right]^2,
\end{equation}
We derive expressions for the noise terms, $P^{\rm noise}(k, \mu)$, in Appendix~\ref{appendix:surveys}. When binned in both $k$ and $\mu$, an annulus of width $(\Delta k, \Delta \mu)$ will contain a number of Fourier modes, $N_{m}$, given by
\begin{equation}
  \label{eq:nmodes}
N_m(k,\mu) = 2 \pi k^2 \Delta k \Delta \mu \left( \frac{V_{\rm survey}}{(2 \pi)^3} \right),
\end{equation}
where $V_{\rm survey}$ is the comoving survey volume (see section \ref{appendix:volume} for further discussion of volume matching).
In practice we are interested in the variance of the spherically-averaged power spectrum, which can be obtained by summing the errors over $\mu$ in inverse quadrature. Hence we can estimate the variance on a given k-mode of the spherically-averaged power spectrum as,
\begin{equation}
\frac{1}{{\rm var}[P_{\rm 21cm \times LAE}(k)]} = \sum_{\mu}
\frac{N_m(k, \mu)}{{\rm var}[P_{\rm 21cm \times LAE}(k, \mu)]} .
\end{equation}
For a given power spectrum we estimate the line-of-sight enhancement due to redshift-space distortions as,
\begin{equation}
P(k, \mu) = (1 + \beta \mu^2)^2 P(k).
\end{equation}
Here $\beta = \Omega^{0.6}_{\rm m} (z)/b$, where $b$ is the bias factor \citep{1987MNRAS.227....1K}
of the respective field. We assume $b_{\rm 21cm} = 1$ and calculate,
\begin{equation}
b^2_{\rm LAE} = P_{\rm LAE}(k) / P_{\rm DM}(k),
\end{equation}
where $P_{\rm DM}(k)$ is the dark matter power spectrum. We employ this linear treatment to avoid the computational overhead of the full radiative transfer calculation (i.e.\@ using the simulation gas peculiar velocities), although note that algorithms for the full treatment do exist \citep[such as][]{2019MNRAS.tmp.2288C}.

\begin{figure*}
  \includegraphics[width=\textwidth]{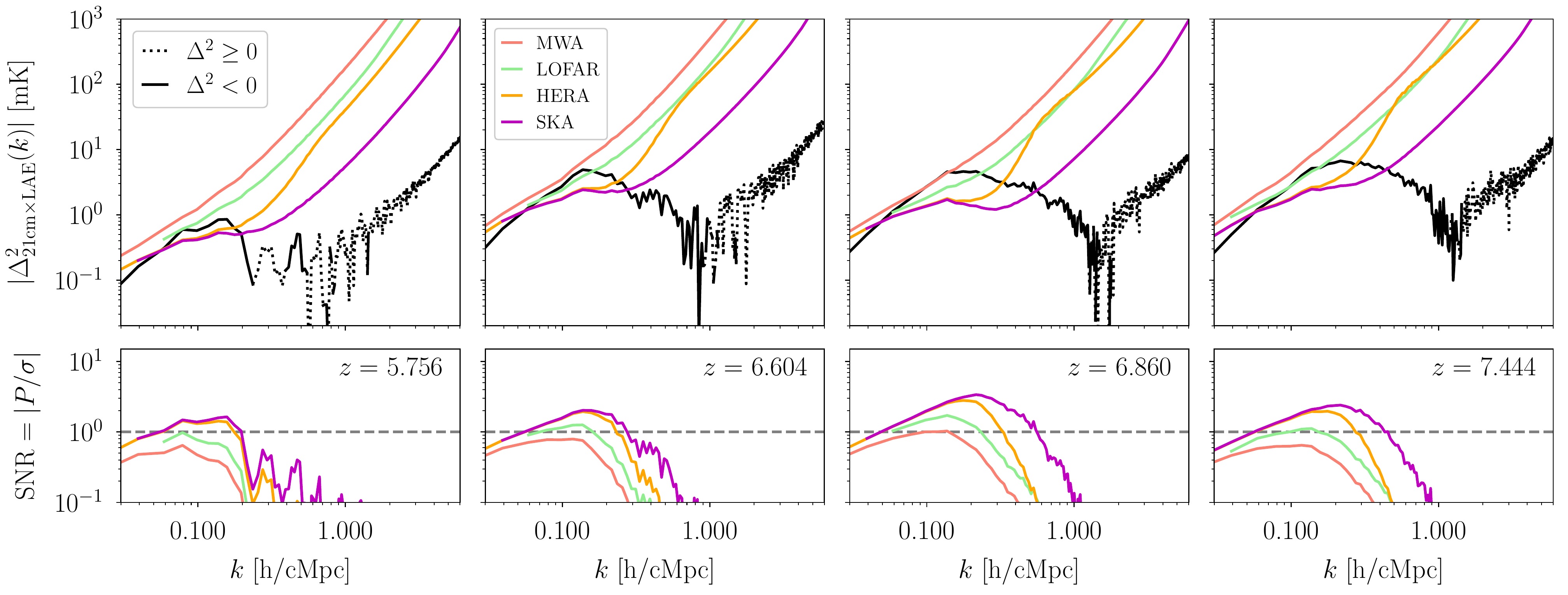}
  \caption{Observational sensitivities for measuring the cross power spectrum of the delayed-end Very Late reionization history. The coloured lines indicate different survey error predictions: red (MWA), green (LOFAR), orange (HERA) and purple (SKA). \emph{Top panels}: The predicted 1-$\sigma$ error measuring the dimensionless cross-power spectrum, with the black line indicating the underlying model power spectrum. As in Figure~\ref{fig:3}, from left to right the panels show increasing redshift, starting at $z=5.756$ on the left and ending with $z=7.444$ on the right. \emph{Bottom panels}: the signal-to-noise ratio for each of the survey sensivity predictions.}
  \label{fig:4}
\end{figure*}

In this work we consider 21-cm observations using the Phase II MWA \citep{2018PASA...35...33W}, LOFAR \citep{2013A&A...556A...2V}, HERA \citep{2017PASP..129d5001D}, and SKA1-low \citep{2016SPIE.9906E..28W} interferometers. For the LAE galaxy survey we estimate the sensitivities of a Subaru survey, using a combination of narrowband selection with the Hyper Suprime-Cam (HSC) \citep{2018PASJ...70S...1M,2017arXiv170407455O} and spectroscopy from the Prime Focus Spectrograph (PFS) \citep{2018SPIE10702E..1CT,2014PASJ...66R...1T}. We assume an LAE survey with the field of view of the HSC Subaru Strategic Program's (SSP) Deep field \citep[which has an area of 27 $\deg^2$,][]{2018PASJ...70S...1M} when calculating LAE sensitivities. Details of the sensitivity modelling of these instruments can be found in Appendix~\ref{appendix:surveys}, and the chosen parameters are shown in Table~\ref{tab:surveys}.

\section{Results}
\label{sec:results}
\subsection{Cross-power spectra}
We plot the spherically averaged cross-power spectra for our six different reionization histories in Figure~\ref{fig:3}, showing the original and delayed-end models in the top and bottom panels, respectively. The different reionization models are shown by the coloured lines with HM12 in blue, Late in purple, and Very Late in orange. The sign of the power spectrum is indicated by the style of the line, with solid representing negative and dotted indicating positive. We note that, as denoted in Eq.~\ref{eq:coeff}, positive values of the cross-power spectrum correspond to correlation between pixels, whereas negative power corresponds to anticorrelation. The shape of the power spectrum seen in most of the panels has a widely peaked negative component at large scales (small $k$), and an approximately power law-shaped positive component at small scales (large $k$).

On large scales we expect the LAEs to be anticorrelated with the 21-cm brightness, as they reside in the more massive overdensities which ionize inside-out \citep{2006MNRAS.369.1625I}. On scales smaller than the typical bubble size we expect to see correlation, with positive power contributions from pairs of pixels in the overdensities around LAEs, where the gas self-shields and must be ionized outside-in \citep{2000ApJ...530....1M}. The general shape of our results is consistent with previous work \citep{2007ApJ...660.1030F,2009ApJ...690..252L,2013MNRAS.432.2615W,2016MNRAS.457..666V,2018MNRAS.479.2754K}

\begin{figure*}
  \includegraphics[width=\textwidth]{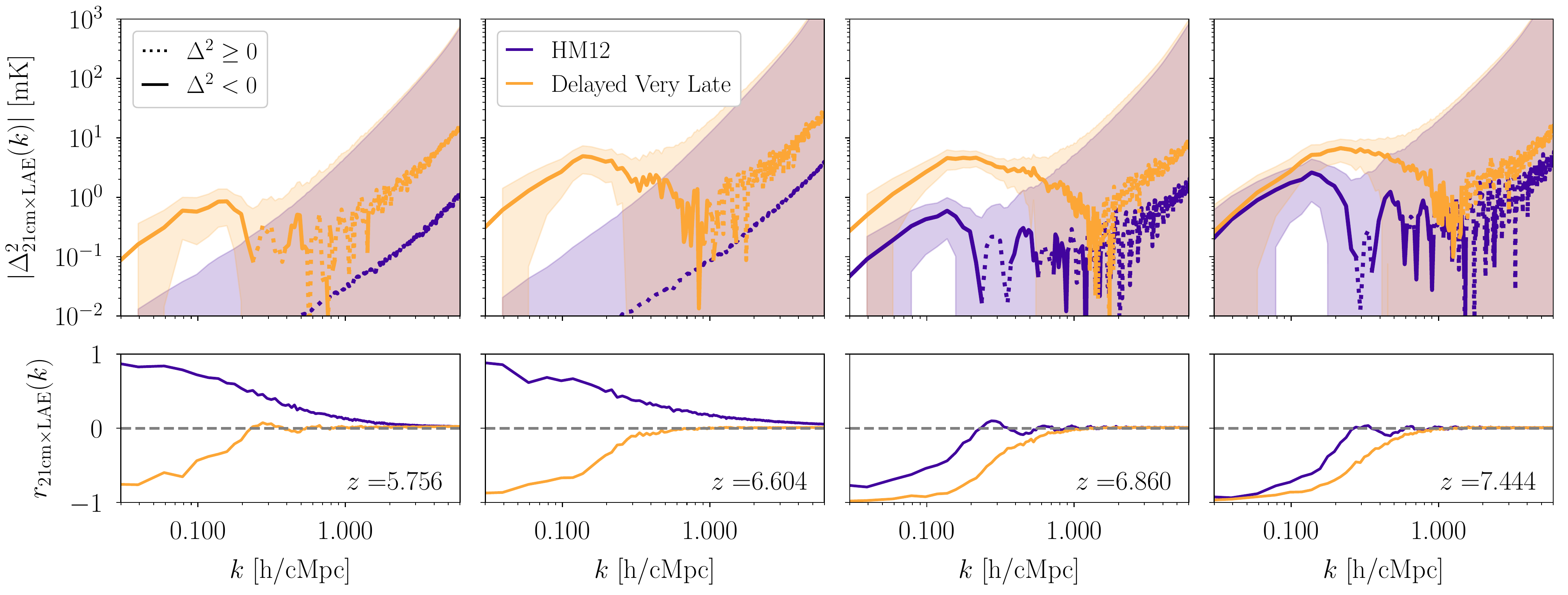}
  \caption{Distinguishing reionization histories using a PFS-SKA survey: a comparison of the HM12 (purple) and delayed-end Very Late (orange) reionization histories. \emph{Top panels:} The cross-power spectrum, with shading indicating the 1-$\sigma$ error predicted for the PFS-SKA survey. \emph{Bottom panels:} the cross-correlation coefficient. As in Figure~\ref{fig:3}, from left to right the panels show increasing redshift, starting at $z=5.756$ on the left and ending with $z=7.444$ on the right.}
  \label{fig:5}
\end{figure*}

In more detail we see that for the original reionization histories --- with reionization ending by $z=6$ --- there is no negative power in the left panel at $z=5.756$. At these redshifts the large-scale IGM is ionized, and so we see only positive correlation resulting from the residual/self-shielded neutral gas in the overdensities that host LAEs. For the delayed-end reionization histories the $z=5.756$ panel shows the hint of the anticorrelation signal at large scales, since there are still neutral islands of gas remaining. At the higher redshifts, for both the original and delayed-end scenarios, we see the anticorrelation signal as described above. Another feature present in many of the power spectra is an oscillatory shape around intermediate scales. This has been observed in other simulations \citep[such as][]{2013MNRAS.432.2615W,2016MNRAS.457..666V}, and represents a harmonic imprinting of the typical bubble size.

Comparing the different reionization histories, we note that the Very Late model has the most power on approximately all scales, and the HM12 model has the least. This is because the amplitude of the power spectrum traces the mean neutral hydrogen fraction, with the Very Late model reionizing latest and hence having the highest mean neutral fraction at a given redshift. We see that the position of the broad peak in the anticorrelation component is dependent on the reionization history, as is the scale when the sign of the power spectrum turns over. This reflects the distribution of bubble sizes, which is a function of the average neutral fraction $\overline{x}_{\rm HI}$. Similarly we see that the small-scale positive correlation component is dependent on the reionization history. This dependency results from the amount of self-shielded neutral gas that is present within ionized regions, controlled by the background photoionization rate $\Gamma_{\rm HI}$. As was similarly found in \citet{2018MNRAS.479.2754K}, we predict slightly more power at small scales than some previous work, which results from a more accurate simulation of the self-shielded gas in the ionized IGM.\@ Our high resolution simulations are better able to resolve the small-scale overdensities, and we employ the self-shielding prescription of \citet{2017arXiv170706993C} to ensure that this effect is properly captured. However we note that our modelling assumes a uniform photoionization rate, which likely acts to enhance the amount of self-shielded gas compared to a more realistic inhomogeneous photoionizing field. At small-scales (large k), the amplitude of the cross-power spectrum is dependent on the self-shielding modelling, and hence is subject to the more modelling uncertainty than the large scales. Whilst the anticorrelation peak in the power spectrum is robust to the self-shielding modelling, we note that the turnover scale depends on the interplay between the ionized bubble distribution and the amount of self-shielding within ionized regions. See Appendix~\ref{appendix:d} for more discussion of the self-shielding assumptions.

For the interested reader, the corresponding correlation functions are provided in Appendix~\ref{appendix:C}.

\subsection{Observational sensitivities}
We plot our predictions for the observational sensitivies of measuring our delayed-end Very Late reonization history in Figure~\ref{fig:4}. The model power spectrum is shown with the black curves (solid and dotted lines indicating negative and positive power, respectively, as in Figure~\ref{fig:3}). The coloured lines show the error predicted for MWA (red), LOFAR (green), HERA (orange) and SKA (purple), combined with a Subaru/PFS LAE survey as detailed in Appendix~\ref{appendix:surveys}. The bottom panels show the predicted signal-to-noise ratios as a function of $k$ for these surveys. We note that the sensitivity predictions are model-dependent; in Figure~\ref{fig:4} we show only the predictions for our delayed-end Very Late model.

The general shape of the sensitivity curves is approximately a power law at large $k$, with a plateau at small $k$. At large $k$, the difference in sensitivity is determined by the different 21-cm survey array configurations. The 21-cm thermal noise on a given line-of-sight mode scales like $\propto \Omega^2/n_b(k_{\bot})$, where $\Omega$ is the field of view of the interferometer, and $n_b$ is the baseline density (see Eq.~\ref{eq:thermal_noise}). Hence, for modes above $k \gtrsim 1$ cMpc/h$^{-1}$, we see that the large density of baselines of the SKA results in the highest sensitivity. At small $k$, on the other hand, the survey sample variance comes into play. Here the different survey volumes limit the number of modes that can be measured. For our chosen survey parameters the limiting volume is the PFS survey rather than the 21-cm surveys (see Appendix \ref{appendix:volume}), and hence all of the lines converge at small $k$. The SKA has a small field of view which results in a smaller survey volume compared to, for example, the MWA; in future cross-correlation surveys limited by the 21-cm survey volume, this will play a role at small $k$.

  Our 21-cm thermal noise predictions are qualitatively consistent with those made by \citet{2016MNRAS.463.2583K}. Our cross-power spectrum error estimates are also broadly consistent with previous work, although we predict somewhat larger errors than those of \citet{2018MNRAS.479.2754K}. This may be due to the different parametrisation of the 21-cm surveys: for example for the SKA we assume a core radius of $r_{\rm core} = 350$ m and an outer radius of $r_{\rm max} = 6.4$ km based on the core array of the proposed V4A configuration (see Table~\ref{tab:surveys} and Appendix \ref{appendix:21cmsurveys}), whilst \citet{2018MNRAS.479.2754K} make estimates for a more compact configuration using $r_{\rm core} = 20$ m and $r_{\rm max} = 1$ km.

\begin{table}
    \caption{Total signal-to-noise ratios for each of the survey combinations measuring the two bracketing reionization histories at $z=6.6$, calculated in the range $0.1 \leq k < 10$ h/cMpc}
    \label{tab:snr}
    \begin{tabular}{l|c|c|c|c}
      Reionization  & MWA & LOFAR & HERA & SKA \\
        \hline
        HM12 & 0.004 & 0.009 & 0.032 & 0.109 \\
        \hline
        Delayed Very Late & 1.427 & 2.494 & 4.495 & 5.531 \\
        \hline
	\end{tabular}
\end{table}

We calculate the total signal-to-noise ratios measured across linearly spaced $k$-bins in quadrature using,
\begin{equation}
{\rm SNR}_{\rm total}^2 = \sum_i {\rm SNR}_i^2 = \sum_i \left( \frac{P(k_i)}{\sigma(k_i)} \right)^2,
\end{equation}
where $i$ indexes the range of $k$-bins between $0.1 \leq k < 10$ h/cMpc. In Table~\ref{tab:snr} we show the total signal-to-noise ratios for the bracketing reionization histories --- the HM12 and Delayed-end Very Late models --- at $z=6.6$, using our sensitivity predictions for the four survey combinations. As expected from Figure~\ref{fig:4}, the signal-to-noise ratio is highest for the PFS-SKA and PFS-HERA combinations. We also note that the Delayed-end Very Late reionization history can be detected in this $k$-range with a total signal-to-noise ratio greater than 5 for the PFS-SKA survey combination.

In Figure~\ref{fig:5} we plot a comparison of the cross-power spectrum evolution for the bracketing HM12 and Delayed-end Very Late reionization histories. We indicate the predicted 1-$\sigma$ errors from a PFS-SKA observational survey using shading. In the bottom panels we show the correlation coefficient for these two models. For the redshifts near the end of reionization we see that these reionization histories could be distinguished at a level of at least 3-sigma (at large scales $k\sim 0.1$ h/cMpc).

\section{Discussion}
\label{sec:discussion}

\subsection{Detecting a delayed end to reionization}
The results in Figures~\ref{fig:4} \&~\ref{fig:5} and Table~\ref{tab:snr} demonstrate the prospects for detecting our ``Delayed-end Very Late'' reionization history in the cross-correlation of 21-cm and LAE observations. The final evolution of the neutral fraction near to the end of reionization has a strong impact on the cross-correlation signal. We have compared our two bracketing reionization histories:
\begin{itemize}
\item \textbf{HM12}, for which reionization has finished by $z=6.7$. This means that at the $z=5.7$ and $z=6.6$ narrowbands, the large-scale IGM is fully ionized and the background photoionization is reaching spatial equilibrium.
  \item \textbf{Delayed-end Very Late}, for which reionization only finishes around $z=5.3$. This means that at the two lower redshift narrowbands there is still significant neutral gas in the IGM.\@
\end{itemize}
We see in Figure~\ref{fig:5} that it is possible to distinguish these different reionization scenarios using a PFS-SKA survey. In particular we note that for the lower redshifts the cross-power spectrum behaves very differently for these scenarios: in the HM12 case the lack of bubble structure leads to only positive power even at large scales, whilst in the delayed-end case we see the imprint of the bubble structure in the form of negative power and a clear turnover scale.

\begin{figure}
  \includegraphics[width=0.5\textwidth]{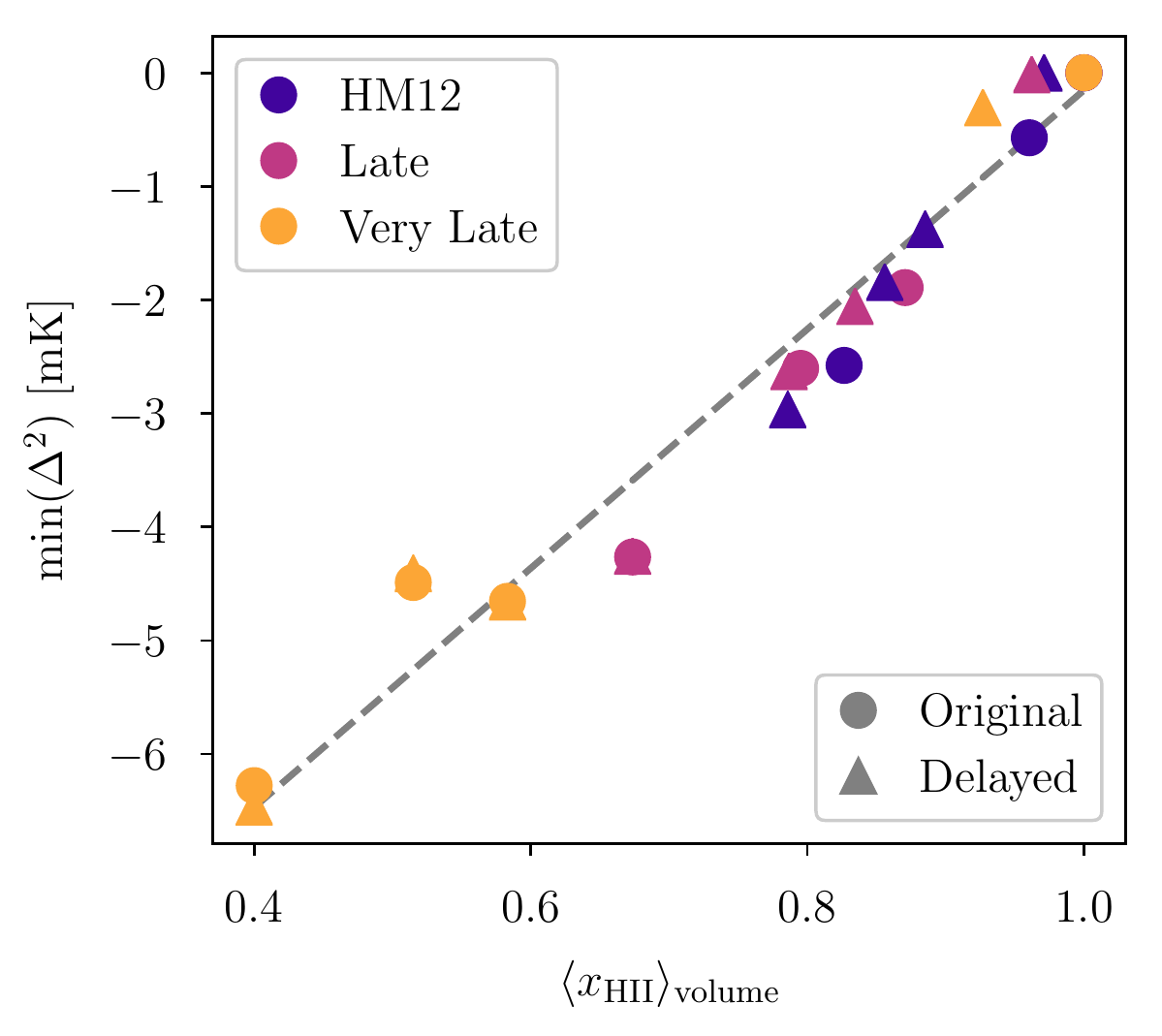}
  \caption{Dependence of the cross-power spectrum features with the reionization history: the correlation between the volume-averaged ionized fraction, $\langle x_{\rm HII} \rangle$, and the size of the negative peak in the power spectrum, $\min(\Delta^2)$. The different points are derived from our six reionization histories, differentiated by the colours and symbols.}
  \label{fig:8}
\end{figure}

We also note in passing that our results suggest it may be possible to detect the signal from ionized bubbles in the cross-power spectrum at redshifts as low as $z=6.6$. As has been noted previously \citep{2019MNRAS.485.3486D,2009ApJ...690..252L,2007ApJ...654...12Z,2004ApJ...613....1F}, whilst reionization is still ongoing the typical bubble size is imprinted on the power spectrum as a turnover in the sign. In our models we find a dip in the power at this turnover scale, which means it may not be possible to separate the power spectrum signal from the noise at that scale. However, we find in all our models (at intermediate redshifts $z\sim 6.6$) that it is possible to detect the power on scales slightly larger than this turnover. For example in the Delayed-end Very Late model, using a PFS-SKA survey at $z=6.6$, it would be possible to measure the power-spectrum at scales around $k\sim 0.2$ h/cMpc, close to the turnover at $k\sim 0.3$ h/cMpc, with a signal-to-noise ratio of $\sim 3$, and hence provide some constraints on the typical bubble size.

\begin{figure}
  \centering
  \includegraphics[width=0.5\textwidth]{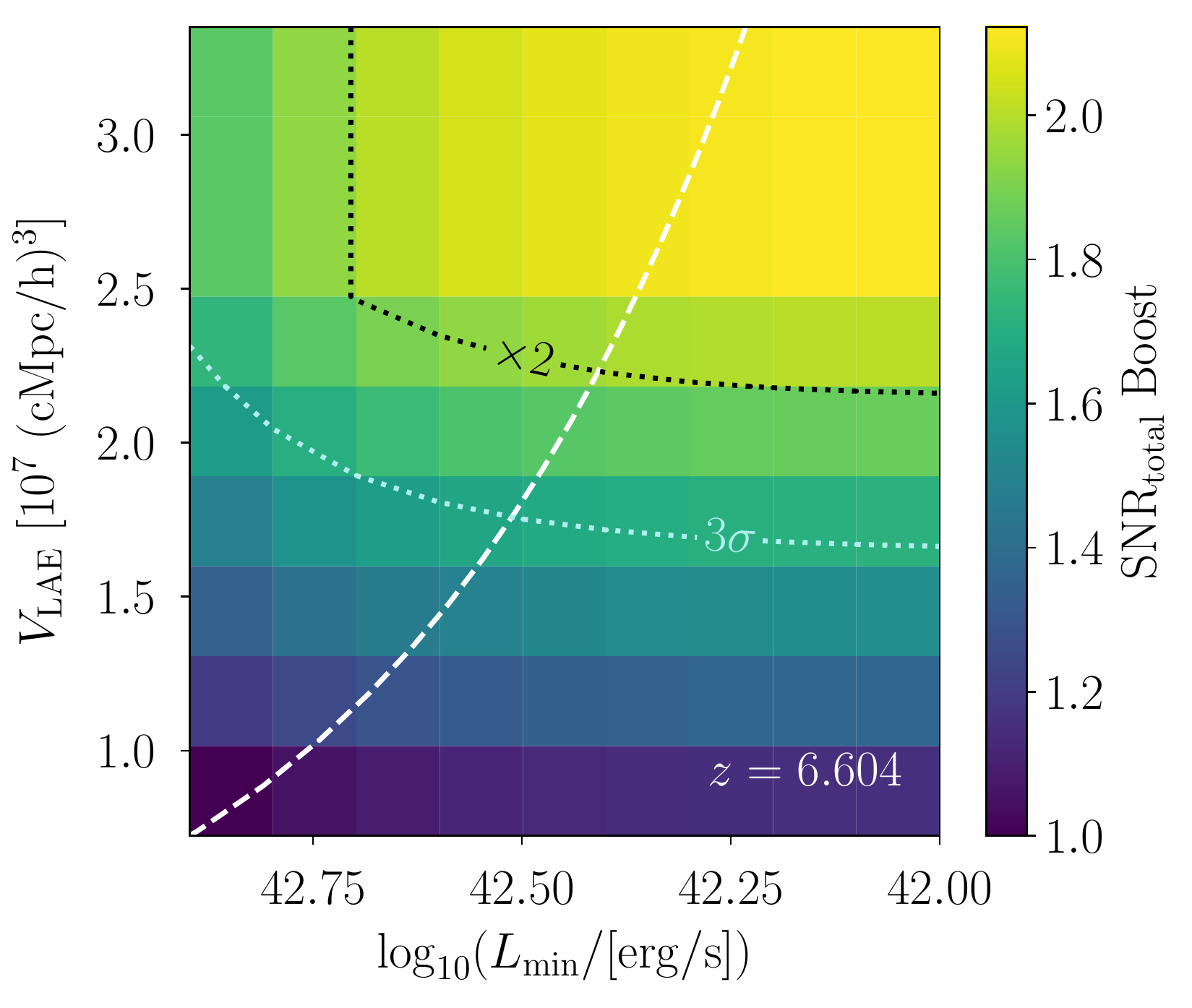}
  \includegraphics[width=0.5\textwidth]{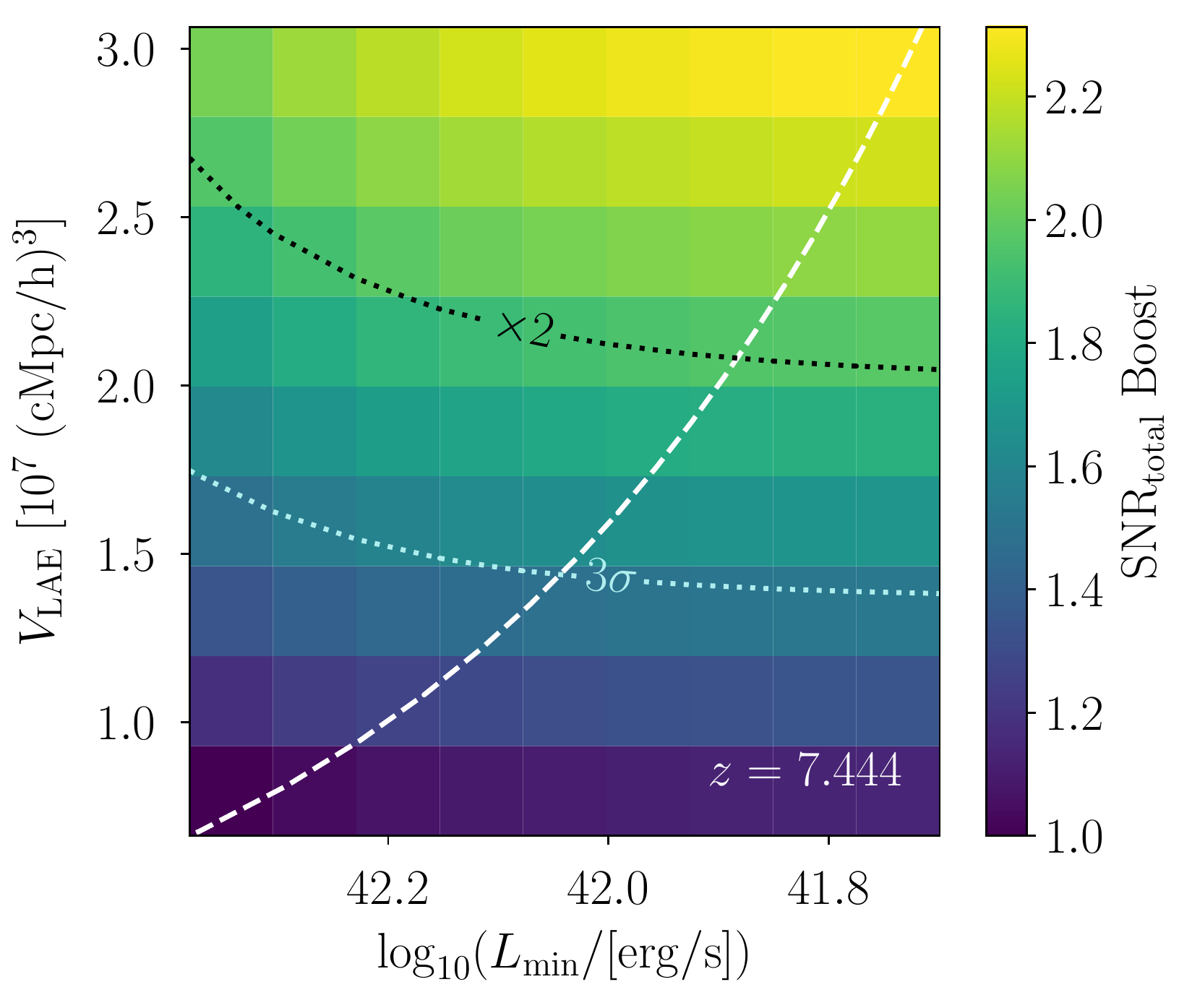}
  \caption{Predictions for the boost in the total signal-to-noise ratio of a PFS-SKA survey measuring the delayed-end Very Late reionization history at $z=6.6$ ($z=7.4$) in the top (bottom) panel, calculated in the range $0.1 \leq k < 10$ h/cMpc. The PFS survey is expanded to have an increased field of view (corresponding to an increased survey volume, $V_{\rm LAE}$, shown on the vertical axis) and depth (parametrised by the minimum observed Ly$\alpha$ luminosity, $L_{\rm min}$, shown on the horizontal axis). The reference SNR$_{\rm total}$ is calculated for our default parameterization of a PFS-SKA survey in which the LAE component of the survey has $A_\theta = 27$ deg$^2$, $V_{\rm LAE} = 2.3 \times 10^7$ cMpc$^3$ ($2.1 \times 10^{7}$ cMpc$^3$) and $L_{\rm min} = 7.9 \times 10^{42}$ erg/s ($2.4 \times 10^{42}$ ergs/s). The black dotted contour indicates the values that give a boost in SNR$_{\rm total}$ of 2, whilst the white dashed line indicates proportional changes in both $V_{\rm LAE}$ and $L_{\rm min}$. The dotted light turquoise line indicates the survey parameters which give enough sensitivity to distinguish the delayed Very Late history from the HM12 history at $k=0.2$ h/cMpc to a 3-$\sigma$ level.}
  \label{fig:9}
\end{figure}

Similarly we note that the size of the negative peak in the cross-power spectrum at large scales may offer another route to constraining the progress of reionization and the typical bubble size. In Figure~\ref{fig:8} we plot the magnitude of the dimensionless cross-power spectrum at this negative peak as a function of the average ionized fraction. We note that this forms a tight correlation, with a more ionized IGM resulting in a smaller peak signal. This reflects the fact that the amplitude of the cross-power spectrum traces the mean neutral fraction. As the position of the turnover scale is dependent on our self-shielding assumptions, the size of the negative peak is possibly a more robust feature for constraining the ionized bubble distribution than the turnover scale. We note however that the sharp transition of the bubble edge in real space is more easily seen in the correlation function (see Appendix~\ref{appendix:C}).

\subsection{Future survey sensitivities}
The results presented in the previous section are predictions for the established Subaru HSC Deep field, which has a field of view of $\sim27$ deg$^2$ and a depth of $r \sim 27$ \citep{2018PASJ...70S...1M,2017arXiv170407455O}. In Figure~\ref{fig:9} we demonstrate the effect of expanding the LAE survey in either area or depth, in combination with an SKA 21-cm survey at $z=6.6$ (top panel) and $z=7.4$ (bottom panel). Increasing the total field of view of the survey reduces the sample variance by allowing more line-of-sight modes to be observed for a given $k$-mode. Decreasing the minimum LAE luminosity (i.e.\@ increasing the depth of the survey) allows the survey to sample larger numbers of galaxies, and hence reduces the shot noise.

We see that a combined factor of three improvement in these quantities\footnote{That is, a threefold increase in area and a threefold decrease in the minimum luminosity.} results in approximately double the total signal-to-noise ratio. We might expect that the survey time scales approximately linearly with changes to the area (at fixed depth), but approximately quadratically with changes to the depth (at fixed area) \citep{2013pss2.book..223D}. Considering the regime of incremental improvements in either area or depth, we see in Figure~\ref{fig:9} that a higher signal-to-noise ratio can be achieved with a smaller increase in observing time by increasing the survey area rather than the depth. When the Subaru/PFS survey volume is comparable to the SKA's survey volume then further improvements in area will have no effect, and hence increasing the depth is then the only option. At higher redshifts such as $z=7.4$, we find that increasing the area of the survey (at fixed depth) provides less of an improvement compared to lower redshifts. For higher redshifts increasing both the depth and the survey area in combination is required to boost the signal-to-noise.

Plans for the WFIRST telescope include proposals for \citep{2015arXiv150303757S}:
\begin{itemize}
\item a High Latitude Survey (HLS) with a field of view of $\sim2000$ deg$^2$ and a depth of 27th magnitude in Y/J/H/F184 bands,
\item the option for deeper sub-fields within the HLS, with fields of view of $\sim20$--50 deg$^2$ and depths of 28--29th magnitude.
\end{itemize}
The WFIRST's Wide Field Instrument will have both imaging and spectroscopy modes. This telescope may therefore offer further prospects for observing the 21cm-LAE cross-correlation beyond Subaru/PFS.\@

\section{Conclusions}
\label{sec:conclusions}
We have employed realistic simulation models of the 21-cm signal and LAE population evolution to predict the 21cm-LAE cross-power spectrum during the epoch of reionization. Our models are calibrated to be consistent with observations of the CMB optical depth, post-reionization Lyman-$\alpha$ forest measurements and LAE population statistics at $z>5$ including the luminosity function, equivalent width distribution and angular clustering function.

We confirm the predictions of recent works \citep{2018MNRAS.479.2754K,2018MNRAS.479L.129H,2017ApJ...836..176H} which suggested that this cross-correlation signal can be observed with future surveys, such as a combination of Subaru/PFS and SKA.\@ We have also found that:
\begin{itemize}
 \item It should be possible to observe the cross-power spectrum signal for a late delayed-end reionization scenario at redshifts as low as $z=5.7$. This scenario has been found to predict the behaviour of the post-reionization Lyman-$\alpha$ forest \citep{2018arXiv180906374K,2019arXiv190512640K,2019arXiv191003570N}, and is consistent with the current Planck contraints on the Thomson optical depth to the CMB \citep{2018arXiv180706209P}. Targeting these lower redshifts with a Subaru/PFS-SKA cross-correlation survey will therefore allow us to constrain the manner in which reionization ended.
 \item It is possible to achieve better sensitivity on the cross-power spectrum by expanding the Subaru HSC survey. Increasing either the total field of view or the depth can reduce the total measurement variance, although for incremental increases we recommend increasing the survey area first. This can provide improved sensitivity until the LAE survey volume is as large as the 21-cm survey volume.
\end{itemize}

The 21cm-LAE cross-correlation signal can provide a robust constraint on the history of reionization, making possible a 21-cm-based inference that might otherwise be plagued by foregrounds. Our simulations suggest that future observations with Subaru/PFS and the SKA --- at redshifts as low as $z=5.7$ --- can provide insight into this final phase transition's end.

\section*{Acknowledgements}
We thank the referee for their helpful comments and suggestions. We also thank Kenji Kubota for fruitful discussions.

LHW is supported by the Science and Technology
Facilities Council (STFC). Support by ERC Advanced Grant 320596 `\emph{The Emergence of Structure
During the Epoch of Reionization}' is gratefully acknowledged.

We acknowledge
PRACE for  awarding  us  access  to  the  Curie  supercomputer,  based in
France  at  the  Tr\'{e}s  Grand  Centre  de  Calcul  (TGCC).
This work was performed using the Cambridge Service for Data Driven Discovery
(CSD3), part of which is operated by the University of Cambridge Research
Computing on behalf of the STFC DiRAC HPC Facility (\url{www.dirac.ac.uk}).
The DiRAC component of CSD3 was funded by BEIS capital funding via STFC
capital grants ST/P002307/1 and ST/R002452/1 and STFC operations
grant ST/R00689X/1. DiRAC is part of the National e-Infrastructure.

The analysis code used in this work was written in
\texttt{Rust} \citep[\url{https://www.rust-lang.org/},][]{Matsakis:2014:RL:2692956.2663188} and
\texttt{Python} \citep[\url{https://www.python.org/},][]{van1995python}. In particular
we employed the \texttt{SciPy}
\citep[\url{https://www.scipy.org/},][]{SCIPY} ecosystem of libraries including: \texttt{NumPy}
\citep[\url{https://www.numpy.org/},][]{NUMPY}, \texttt{Matplotlib}
\citep[\url{https://matplotlib.org/},][]{MATPLOTLIB} and \texttt{Cython}
\citep[\url{https://cython.org/},][]{CYTHON}. We also made use of the
\texttt{hankel} \citep[\url{https://github.com/steven-murray/hankel},][]{HANKEL} package.




\bibliographystyle{mnras}
\bibliography{references.bib}

\begin{thebibliography}{}
\makeatletter
\relax
\def\mn@urlcharsother{\let\do\@makeother \do\$\do\&\do\#\do\^\do\_\do\%\do\~}
\def\mn@doi{\begingroup\mn@urlcharsother \@ifnextchar [ {\mn@doi@}
  {\mn@doi@[]}}
\def\mn@doi@[#1]#2{\def\@tempa{#1}\ifx\@tempa\@empty \href
  {http://dx.doi.org/#2} {doi:#2}\else \href {http://dx.doi.org/#2} {#1}\fi
  \endgroup}
\def\mn@eprint#1#2{\mn@eprint@#1:#2::\@nil}
\def\mn@eprint@arXiv#1{\href {http://arxiv.org/abs/#1} {{\tt arXiv:#1}}}
\def\mn@eprint@dblp#1{\href {http://dblp.uni-trier.de/rec/bibtex/#1.xml}
  {dblp:#1}}
\def\mn@eprint@#1:#2:#3:#4\@nil{\def\@tempa {#1}\def\@tempb {#2}\def\@tempc
  {#3}\ifx \@tempc \@empty \let \@tempc \@tempb \let \@tempb \@tempa \fi \ifx
  \@tempb \@empty \def\@tempb {arXiv}\fi \@ifundefined
  {mn@eprint@\@tempb}{\@tempb:\@tempc}{\expandafter \expandafter \csname
  mn@eprint@\@tempb\endcsname \expandafter{\@tempc}}}

\bibitem[\protect\citeauthoryear{{Barkana} \& {Loeb}}{{Barkana} \&
  {Loeb}}{2001}]{2001PhR...349..125B}
{Barkana} R.,  {Loeb} A.,  2001, \mn@doi [\physrep]
  {10.1016/S0370-1573(01)00019-9}, \href
  {http://adsabs.harvard.edu/abs/2001PhR...349..125B} {349, 125}

\bibitem[\protect\citeauthoryear{{Beane} \& {Lidz}}{{Beane} \&
  {Lidz}}{2018}]{2018ApJ...867...26B}
{Beane} A.,  {Lidz} A.,  2018, \mn@doi [\apj] {10.3847/1538-4357/aae388}, \href
  {https://ui.adsabs.harvard.edu/abs/2018ApJ...867...26B} {867, 26}

\bibitem[\protect\citeauthoryear{{Becker}, {Bolton}, {Madau}, {Pettini},
  {Ryan-Weber}  \& {Venemans}}{{Becker} et~al.}{2015}]{2015MNRAS.447.3402B}
{Becker} G.~D.,  {Bolton} J.~S.,  {Madau} P.,  {Pettini} M.,  {Ryan-Weber}
  E.~V.,   {Venemans} B.~P.,  2015, \mn@doi [\mnras] {10.1093/mnras/stu2646},
  \href {http://adsabs.harvard.edu/abs/2015MNRAS.447.3402B} {447, 3402}

\bibitem[\protect\citeauthoryear{Behnel, Bradshaw, Citro, Dalcin, Seljebotn  \&
  Smith}{Behnel et~al.}{2011}]{CYTHON}
Behnel S.,  Bradshaw R.,  Citro C.,  Dalcin L.,  Seljebotn D.~S.,   Smith K.,
  2011, \mn@doi [Computing in Science Engineering] {10.1109/MCSE.2010.118}, 13,
  31

\bibitem[\protect\citeauthoryear{{Bolton}, {Puchwein}, {Sijacki}, {Haehnelt},
  {Kim}, {Meiksin}, {Regan}  \& {Viel}}{{Bolton}
  et~al.}{2017}]{2017MNRAS.464..897B}
{Bolton} J.~S.,  {Puchwein} E.,  {Sijacki} D.,  {Haehnelt} M.~G.,  {Kim} T.-S.,
   {Meiksin} A.,  {Regan} J.~A.,   {Viel} M.,  2017, \mn@doi [\mnras]
  {10.1093/mnras/stw2397}, \href
  {http://adsabs.harvard.edu/abs/2017MNRAS.464..897B} {464, 897}

\bibitem[\protect\citeauthoryear{{Chang}, {Gong}, {Santos}, {Silva}, {Aguirre},
  {Dor{\'e}}  \& {Pritchard}}{{Chang} et~al.}{2015}]{2015aska.confE...4C}
{Chang} T.~C.,  {Gong} Y.,  {Santos} M.,  {Silva} M.~B.,  {Aguirre} J.,
  {Dor{\'e}} O.,   {Pritchard} J.,  2015, in Advancing Astrophysics with the
  Square Kilometre Array (AASKA14). p.~4 (\mn@eprint {arXiv} {1501.04654})

\bibitem[\protect\citeauthoryear{{Chapman} \& {Santos}}{{Chapman} \&
  {Santos}}{2019}]{2019MNRAS.tmp.2288C}
{Chapman} E.,  {Santos} M.~G.,  2019, \mn@doi [\mnras] {10.1093/mnras/stz2663},
  \href {https://ui.adsabs.harvard.edu/abs/2019MNRAS.tmp.2288C} {p.~2288}

\bibitem[\protect\citeauthoryear{{Chapman} et~al.,}{{Chapman}
  et~al.}{2015}]{2015aska.confE...5C}
{Chapman} E.,  et~al., 2015, in Advancing Astrophysics with the Square
  Kilometre Array (AASKA14). p.~5 (\mn@eprint {arXiv} {1501.04429})

\bibitem[\protect\citeauthoryear{{Chardin}, {Haehnelt}, {Aubert}  \&
  {Puchwein}}{{Chardin} et~al.}{2015}]{2015MNRAS.453.2943C}
{Chardin} J.,  {Haehnelt} M.~G.,  {Aubert} D.,   {Puchwein} E.,  2015, \mn@doi
  [\mnras] {10.1093/mnras/stv1786}, \href
  {http://adsabs.harvard.edu/abs/2015MNRAS.453.2943C} {453, 2943}

\bibitem[\protect\citeauthoryear{{Chardin}, {Kulkarni}  \&
  {Haehnelt}}{{Chardin} et~al.}{2018}]{2017arXiv170706993C}
{Chardin} J.,  {Kulkarni} G.,   {Haehnelt} M.~G.,  2018, \mn@doi [\mnras]
  {10.1093/mnras/sty992}, \href
  {http://adsabs.harvard.edu/abs/2018MNRAS.tmp.1018C} {}

\bibitem[\protect\citeauthoryear{{Choudhury}}{{Choudhury}}{2009}]{2009CSci...97..841C}
{Choudhury} T.~R.,  2009, Current Science, \href
  {http://adsabs.harvard.edu/abs/2009CSci...97..841C} {97, 841}

\bibitem[\protect\citeauthoryear{{Choudhury}, {Haehnelt}  \&
  {Regan}}{{Choudhury} et~al.}{2009}]{2009MNRAS.394..960C}
{Choudhury} T.~R.,  {Haehnelt} M.~G.,   {Regan} J.,  2009, \mn@doi [\mnras]
  {10.1111/j.1365-2966.2008.14383.x}, \href
  {http://adsabs.harvard.edu/abs/2009MNRAS.394..960C} {394, 960}

\bibitem[\protect\citeauthoryear{{Choudhury}, {Puchwein}, {Haehnelt}  \&
  {Bolton}}{{Choudhury} et~al.}{2015}]{2015MNRAS.452..261C}
{Choudhury} T.~R.,  {Puchwein} E.,  {Haehnelt} M.~G.,   {Bolton} J.~S.,  2015,
  \mn@doi [\mnras] {10.1093/mnras/stv1250}, \href
  {http://adsabs.harvard.edu/abs/2015MNRAS.452..261C} {452, 261}

\bibitem[\protect\citeauthoryear{{Datta}, {Bharadwaj}  \& {Choudhury}}{{Datta}
  et~al.}{2007}]{2007MNRAS.382..809D}
{Datta} K.~K.,  {Bharadwaj} S.,   {Choudhury} T.~R.,  2007, \mn@doi [\mnras]
  {10.1111/j.1365-2966.2007.12421.x}, \href
  {https://ui.adsabs.harvard.edu/abs/2007MNRAS.382..809D} {382, 809}

\bibitem[\protect\citeauthoryear{{Dayal} \& {Ferrara}}{{Dayal} \&
  {Ferrara}}{2018}]{2018PhR...780....1D}
{Dayal} P.,  {Ferrara} A.,  2018, \mn@doi [\physrep]
  {10.1016/j.physrep.2018.10.002}, \href
  {https://ui.adsabs.harvard.edu/abs/2018PhR...780....1D} {780, 1}

\bibitem[\protect\citeauthoryear{{DeBoer} et~al.,}{{DeBoer}
  et~al.}{2017}]{2017PASP..129d5001D}
{DeBoer} D.~R.,  et~al., 2017, \mn@doi [\pasp]
  {10.1088/1538-3873/129/974/045001}, \href
  {https://ui.adsabs.harvard.edu/abs/2017PASP..129d5001D} {129, 045001}

\bibitem[\protect\citeauthoryear{{Dijkstra}}{{Dijkstra}}{2014}]{2014PASA...31...40D}
{Dijkstra} M.,  2014, \mn@doi [\pasa] {10.1017/pasa.2014.33}, \href
  {http://adsabs.harvard.edu/abs/2014PASA...31...40D} {31, e040}

\bibitem[\protect\citeauthoryear{{Dijkstra} \& {Wyithe}}{{Dijkstra} \&
  {Wyithe}}{2012}]{2012MNRAS.419.3181D}
{Dijkstra} M.,  {Wyithe} J. S.~B.,  2012, \mn@doi [\mnras]
  {10.1111/j.1365-2966.2011.19958.x}, \href
  {https://ui.adsabs.harvard.edu/#abs/2012MNRAS.419.3181D} {419, 3181}

\bibitem[\protect\citeauthoryear{{Dijkstra}, {Lidz}  \& {Wyithe}}{{Dijkstra}
  et~al.}{2007}]{2007MNRAS.377.1175D}
{Dijkstra} M.,  {Lidz} A.,   {Wyithe} J.~S.~B.,  2007, \mn@doi [\mnras]
  {10.1111/j.1365-2966.2007.11666.x}, \href
  {http://adsabs.harvard.edu/abs/2007MNRAS.377.1175D} {377, 1175}

\bibitem[\protect\citeauthoryear{{Djorgovski}, {Mahabal}, {Drake}, {Graham}  \&
  {Donalek}}{{Djorgovski} et~al.}{2013}]{2013pss2.book..223D}
{Djorgovski} S.~G.,  {Mahabal} A.,  {Drake} A.,  {Graham} M.,   {Donalek} C.,
  2013, {Sky Surveys}.
p.~223, \mn@doi{10.1007/978-94-007-5618-2_5}

\bibitem[\protect\citeauthoryear{{Dumitru}, {Kulkarni}, {Lagache}  \&
  {Haehnelt}}{{Dumitru} et~al.}{2019}]{2019MNRAS.485.3486D}
{Dumitru} S.,  {Kulkarni} G.,  {Lagache} G.,   {Haehnelt} M.~G.,  2019, \mn@doi
  [\mnras] {10.1093/mnras/stz617}, \href
  {https://ui.adsabs.harvard.edu/abs/2019MNRAS.485.3486D} {485, 3486}

\bibitem[\protect\citeauthoryear{{Duncan} \& {Conselice}}{{Duncan} \&
  {Conselice}}{2015}]{2015MNRAS.451.2030D}
{Duncan} K.,  {Conselice} C.~J.,  2015, \mn@doi [\mnras]
  {10.1093/mnras/stv1049}, \href
  {http://adsabs.harvard.edu/abs/2015MNRAS.451.2030D} {451, 2030}

\bibitem[\protect\citeauthoryear{{Efstathiou}}{{Efstathiou}}{1992}]{1992MNRAS.256P..43E}
{Efstathiou} G.,  1992, \mn@doi [\mnras] {10.1093/mnras/256.1.43P}, \href
  {https://ui.adsabs.harvard.edu/abs/1992MNRAS.256P..43E} {256, 43P}

\bibitem[\protect\citeauthoryear{{Feldman}, {Kaiser}  \& {Peacock}}{{Feldman}
  et~al.}{1994}]{1994ApJ...426...23F}
{Feldman} H.~A.,  {Kaiser} N.,   {Peacock} J.~A.,  1994, \mn@doi [\apj]
  {10.1086/174036}, \href
  {https://ui.adsabs.harvard.edu/abs/1994ApJ...426...23F} {426, 23}

\bibitem[\protect\citeauthoryear{{Feng}, {Cooray}  \& {Keating}}{{Feng}
  et~al.}{2017}]{2017arXiv170107005F}
{Feng} C.,  {Cooray} A.,   {Keating} B.,  2017, preprint, \href
  {http://adsabs.harvard.edu/abs/2017arXiv170107005F} {} (\mn@eprint {arXiv}
  {1701.07005})

\bibitem[\protect\citeauthoryear{Frigo \& Johnson}{Frigo \&
  Johnson}{2005}]{FFTW05}
Frigo M.,  Johnson S.~G.,  2005, Proceedings of the IEEE, 93, 216

\bibitem[\protect\citeauthoryear{{Furlanetto} \& {Lidz}}{{Furlanetto} \&
  {Lidz}}{2007}]{2007ApJ...660.1030F}
{Furlanetto} S.~R.,  {Lidz} A.,  2007, \mn@doi [\apj] {10.1086/513009}, \href
  {https://ui.adsabs.harvard.edu/abs/2007ApJ...660.1030F} {660, 1030}

\bibitem[\protect\citeauthoryear{{Furlanetto}, {Zaldarriaga}  \&
  {Hernquist}}{{Furlanetto} et~al.}{2004}]{2004ApJ...613....1F}
{Furlanetto} S.~R.,  {Zaldarriaga} M.,   {Hernquist} L.,  2004, \mn@doi [\apj]
  {10.1086/423025}, \href {http://adsabs.harvard.edu/abs/2004ApJ...613....1F}
  {613, 1}

\bibitem[\protect\citeauthoryear{{Furlanetto}, {Oh}  \& {Briggs}}{{Furlanetto}
  et~al.}{2006}]{2006PhR...433..181F}
{Furlanetto} S.~R.,  {Oh} S.~P.,   {Briggs} F.~H.,  2006, \mn@doi [\physrep]
  {10.1016/j.physrep.2006.08.002}, \href
  {http://adsabs.harvard.edu/abs/2006PhR...433..181F} {433, 181}

\bibitem[\protect\citeauthoryear{{Geil}, {Gaensler}  \& {Wyithe}}{{Geil}
  et~al.}{2011}]{2011MNRAS.418..516G}
{Geil} P.~M.,  {Gaensler} B.~M.,   {Wyithe} J. S.~B.,  2011, \mn@doi [\mnras]
  {10.1111/j.1365-2966.2011.19509.x}, \href
  {https://ui.adsabs.harvard.edu/abs/2011MNRAS.418..516G} {418, 516}

\bibitem[\protect\citeauthoryear{{Giri}, {Mellema}, {Dixon}  \& {Iliev}}{{Giri}
  et~al.}{2018}]{2018MNRAS.473.2949G}
{Giri} S.~K.,  {Mellema} G.,  {Dixon} K.~L.,   {Iliev} I.~T.,  2018, \mn@doi
  [\mnras] {10.1093/mnras/stx2539}, \href
  {https://ui.adsabs.harvard.edu/#abs/2018MNRAS.473.2949G} {473, 2949}

\bibitem[\protect\citeauthoryear{{Gronke}, {Bull}  \& {Dijkstra}}{{Gronke}
  et~al.}{2015}]{2015ApJ...812..123G}
{Gronke} M.,  {Bull} P.,   {Dijkstra} M.,  2015, \mn@doi [\apj]
  {10.1088/0004-637X/812/2/123}, \href
  {http://adsabs.harvard.edu/abs/2015ApJ...812..123G} {812, 123}

\bibitem[\protect\citeauthoryear{{Gunn} \& {Peterson}}{{Gunn} \&
  {Peterson}}{1965}]{1965ApJ...142.1633G}
{Gunn} J.~E.,  {Peterson} B.~A.,  1965, \mn@doi [\apj] {10.1086/148444}, \href
  {https://ui.adsabs.harvard.edu/#abs/1965ApJ...142.1633G} {142, 1633}

\bibitem[\protect\citeauthoryear{{Haardt} \& {Madau}}{{Haardt} \&
  {Madau}}{2012}]{2012ApJ...746..125H}
{Haardt} F.,  {Madau} P.,  2012, \mn@doi [\apj] {10.1088/0004-637X/746/2/125},
  \href {http://adsabs.harvard.edu/abs/2012ApJ...746..125H} {746, 125}

\bibitem[\protect\citeauthoryear{{Hassan}, {Dav{\'e}}, {Finlator}  \&
  {Santos}}{{Hassan} et~al.}{2016}]{2016MNRAS.457.1550H}
{Hassan} S.,  {Dav{\'e}} R.,  {Finlator} K.,   {Santos} M.~G.,  2016, \mn@doi
  [\mnras] {10.1093/mnras/stv3001}, \href
  {http://adsabs.harvard.edu/abs/2016MNRAS.457.1550H} {457, 1550}

\bibitem[\protect\citeauthoryear{Hunter}{Hunter}{2007}]{MATPLOTLIB}
Hunter J.~D.,  2007, \mn@doi [Computing in Science Engineering]
  {10.1109/MCSE.2007.55}, 9, 90

\bibitem[\protect\citeauthoryear{{Hutter}, {Dayal}, {Partl}  \&
  {M{\"u}ller}}{{Hutter} et~al.}{2014}]{2014MNRAS.441.2861H}
{Hutter} A.,  {Dayal} P.,  {Partl} A.~M.,   {M{\"u}ller} V.,  2014, \mn@doi
  [\mnras] {10.1093/mnras/stu791}, \href
  {https://ui.adsabs.harvard.edu/#abs/2014MNRAS.441.2861H} {441, 2861}

\bibitem[\protect\citeauthoryear{{Hutter}, {Dayal}  \& {M{\"u}ller}}{{Hutter}
  et~al.}{2015}]{2015MNRAS.450.4025H}
{Hutter} A.,  {Dayal} P.,   {M{\"u}ller} V.,  2015, \mn@doi [\mnras]
  {10.1093/mnras/stv929}, \href
  {https://ui.adsabs.harvard.edu/#abs/2015MNRAS.450.4025H} {450, 4025}

\bibitem[\protect\citeauthoryear{{Hutter}, {Dayal}, {M{\"u}ller}  \&
  {Trott}}{{Hutter} et~al.}{2017}]{2017ApJ...836..176H}
{Hutter} A.,  {Dayal} P.,  {M{\"u}ller} V.,   {Trott} C.~M.,  2017, \mn@doi
  [\apj] {10.3847/1538-4357/836/2/176}, \href
  {https://ui.adsabs.harvard.edu/abs/2017ApJ...836..176H} {836, 176}

\bibitem[\protect\citeauthoryear{{Hutter}, {Trott}  \& {Dayal}}{{Hutter}
  et~al.}{2018}]{2018MNRAS.479L.129H}
{Hutter} A.,  {Trott} C.~M.,   {Dayal} P.,  2018, \mn@doi [\mnras]
  {10.1093/mnrasl/sly115}, \href
  {https://ui.adsabs.harvard.edu/abs/2018MNRAS.479L.129H} {479, L129}

\bibitem[\protect\citeauthoryear{{Iliev}, {Mellema}, {Pen}, {Merz}, {Shapiro}
  \& {Alvarez}}{{Iliev} et~al.}{2006}]{2006MNRAS.369.1625I}
{Iliev} I.~T.,  {Mellema} G.,  {Pen} U.~L.,  {Merz} H.,  {Shapiro} P.~R.,
  {Alvarez} M.~A.,  2006, \mn@doi [Monthly Notices of the Royal Astronomical
  Society] {10.1111/j.1365-2966.2006.10502.x}, \href
  {https://ui.adsabs.harvard.edu/abs/2006MNRAS.369.1625I} {369, 1625}

\bibitem[\protect\citeauthoryear{{Inoue} et~al.,}{{Inoue}
  et~al.}{2018}]{2018arXiv180100067I}
{Inoue} A.~K.,  et~al., 2018, \mn@doi [\pasj] {10.1093/pasj/psy048}, \href
  {http://adsabs.harvard.edu/abs/2018PASJ..tmp...62I} {}

\bibitem[\protect\citeauthoryear{{Itoh} et~al.,}{{Itoh}
  et~al.}{2018}]{2018arXiv180505944I}
{Itoh} R.,  et~al., 2018, preprint, \href
  {http://adsabs.harvard.edu/abs/2018arXiv180505944I} {} (\mn@eprint {arXiv}
  {1805.05944})

\bibitem[\protect\citeauthoryear{{Jensen}, {Laursen}, {Mellema}, {Iliev},
  {Sommer-Larsen}  \& {Shapiro}}{{Jensen} et~al.}{2013}]{2013MNRAS.428.1366J}
{Jensen} H.,  {Laursen} P.,  {Mellema} G.,  {Iliev} I.~T.,  {Sommer-Larsen} J.,
    {Shapiro} P.~R.,  2013, \mn@doi [\mnras] {10.1093/mnras/sts116}, \href
  {https://ui.adsabs.harvard.edu/\#abs/2013MNRAS.428.1366J} {428, 1366}

\bibitem[\protect\citeauthoryear{{Jones}, {Oliphant}, {Peterson}
  et~al.}{{Jones} et~al.}{2001}]{SCIPY}
{Jones} E.,  {Oliphant} T.,  {Peterson} P.,   et~al., 2001, {SciPy}: Open
  source scientific tools for {Python}, \url {http://www.scipy.org/}

\bibitem[\protect\citeauthoryear{{Kaiser}}{{Kaiser}}{1987}]{1987MNRAS.227....1K}
{Kaiser} N.,  1987, \mn@doi [\mnras] {10.1093/mnras/227.1.1}, \href
  {https://ui.adsabs.harvard.edu/abs/1987MNRAS.227....1K} {227, 1}

\bibitem[\protect\citeauthoryear{{Kakiichi}, {Dijkstra}, {Ciardi}  \&
  {Graziani}}{{Kakiichi} et~al.}{2016}]{2016MNRAS.463.4019K}
{Kakiichi} K.,  {Dijkstra} M.,  {Ciardi} B.,   {Graziani} L.,  2016, \mn@doi
  [\mnras] {10.1093/mnras/stw2193}, \href
  {http://adsabs.harvard.edu/abs/2016MNRAS.463.4019K} {463, 4019}

\bibitem[\protect\citeauthoryear{{Katz} et~al.,}{{Katz}
  et~al.}{2019a}]{2019arXiv190511414K}
{Katz} H.,  et~al., 2019a, arXiv e-prints, \href
  {https://ui.adsabs.harvard.edu/abs/2019arXiv190511414K} {p. arXiv:1905.11414}

\bibitem[\protect\citeauthoryear{{Katz}, {Kimm}, {Haehnelt}, {Sijacki},
  {Rosdahl}  \& {Blaizot}}{{Katz} et~al.}{2019b}]{2019MNRAS.483.1029K}
{Katz} H.,  {Kimm} T.,  {Haehnelt} M.~G.,  {Sijacki} D.,  {Rosdahl} J.,
  {Blaizot} J.,  2019b, \mn@doi [Monthly Notices of the Royal Astronomical
  Society] {10.1093/mnras/sty3154}, \href
  {https://ui.adsabs.harvard.edu/abs/2019MNRAS.483.1029K} {483, 1029}

\bibitem[\protect\citeauthoryear{{Keating}, {Weinberger}, {Kulkarni},
  {Haehnelt}, {Chardin}  \& {Aubert}}{{Keating}
  et~al.}{2020}]{2019arXiv190512640K}
{Keating} L.~C.,  {Weinberger} L.~H.,  {Kulkarni} G.,  {Haehnelt} M.~G.,
  {Chardin} J.,   {Aubert} D.,  2020, \mn@doi [\mnras] {10.1093/mnras/stz3083},
  \href {https://ui.adsabs.harvard.edu/abs/2020MNRAS.491.1736K} {491, 1736}

\bibitem[\protect\citeauthoryear{{Konno} et~al.,}{{Konno}
  et~al.}{2014}]{2014ApJ...797...16K}
{Konno} A.,  et~al., 2014, \mn@doi [\apj] {10.1088/0004-637X/797/1/16}, \href
  {http://adsabs.harvard.edu/abs/2014ApJ...797...16K} {797, 16}

\bibitem[\protect\citeauthoryear{{Konno} et~al.,}{{Konno}
  et~al.}{2018}]{2017arXiv170501222K}
{Konno} A.,  et~al., 2018, \mn@doi [\pasj] {10.1093/pasj/psx131}, \href
  {http://adsabs.harvard.edu/abs/2018PASJ...70S..16K} {70, S16}

\bibitem[\protect\citeauthoryear{{Kubota}, {Yoshiura}, {Takahashi}, {Hasegawa},
  {Yajima}, {Ouchi}, {Pindor}  \& {Webster}}{{Kubota}
  et~al.}{2018}]{2018MNRAS.479.2754K}
{Kubota} K.,  {Yoshiura} S.,  {Takahashi} K.,  {Hasegawa} K.,  {Yajima} H.,
  {Ouchi} M.,  {Pindor} B.,   {Webster} R.~L.,  2018, \mn@doi [\mnras]
  {10.1093/mnras/sty1471}, \href
  {https://ui.adsabs.harvard.edu/abs/2018MNRAS.479.2754K} {479, 2754}

\bibitem[\protect\citeauthoryear{{Kubota}, {Inoue}, {Hasegawa}  \&
  {Takahashi}}{{Kubota} et~al.}{2019}]{2019arXiv191002361K}
{Kubota} K.,  {Inoue} A.~K.,  {Hasegawa} K.,   {Takahashi} K.,  2019, arXiv
  e-prints, \href {https://ui.adsabs.harvard.edu/abs/2019arXiv191002361K} {p.
  arXiv:1910.02361}

\bibitem[\protect\citeauthoryear{{Kulkarni}, {Choudhury}, {Puchwein}  \&
  {Haehnelt}}{{Kulkarni} et~al.}{2016}]{2016MNRAS.463.2583K}
{Kulkarni} G.,  {Choudhury} T.~R.,  {Puchwein} E.,   {Haehnelt} M.~G.,  2016,
  \mn@doi [\mnras] {10.1093/mnras/stw2168}, \href
  {http://adsabs.harvard.edu/abs/2016MNRAS.463.2583K} {463, 2583}

\bibitem[\protect\citeauthoryear{{Kulkarni}, {Choudhury}, {Puchwein}  \&
  {Haehnelt}}{{Kulkarni} et~al.}{2017}]{2017arXiv170104408K}
{Kulkarni} G.,  {Choudhury} T.~R.,  {Puchwein} E.,   {Haehnelt} M.~G.,  2017,
  \mn@doi [\mnras] {10.1093/mnras/stx1167}, \href
  {http://adsabs.harvard.edu/abs/2017MNRAS.469.4283K} {469, 4283}

\bibitem[\protect\citeauthoryear{{Kulkarni}, {Keating}, {Haehnelt}, {Bosman},
  {Puchwein}, {Chardin}  \& {Aubert}}{{Kulkarni}
  et~al.}{2019}]{2018arXiv180906374K}
{Kulkarni} G.,  {Keating} L.~C.,  {Haehnelt} M.~G.,  {Bosman} S. E.~I.,
  {Puchwein} E.,  {Chardin} J.,   {Aubert} D.,  2019, \mn@doi [\mnras]
  {10.1093/mnrasl/slz025}, \href
  {https://ui.adsabs.harvard.edu/abs/2019MNRAS.485L..24K} {485, L24}

\bibitem[\protect\citeauthoryear{{Laursen}, {Sommer-Larsen}  \&
  {Razoumov}}{{Laursen} et~al.}{2011}]{2011ApJ...728...52L}
{Laursen} P.,  {Sommer-Larsen} J.,   {Razoumov} A.~O.,  2011, \mn@doi [\apj]
  {10.1088/0004-637X/728/1/52}, \href
  {https://ui.adsabs.harvard.edu/#abs/2011ApJ...728...52L} {728, 52}

\bibitem[\protect\citeauthoryear{{Laursen}, {Sommer-Larsen}, {Milvang-Jensen},
  {Fynbo}  \& {Razoumov}}{{Laursen} et~al.}{2018}]{2018arXiv180607392L}
{Laursen} P.,  {Sommer-Larsen} J.,  {Milvang-Jensen} B.,  {Fynbo} J. P.~U.,
  {Razoumov} A.~O.,  2018, preprint, \href
  {https://ui.adsabs.harvard.edu/#abs/2018arXiv180607392L} {p.
  arXiv:1806.07392} (\mn@eprint {arXiv} {1806.07392})

\bibitem[\protect\citeauthoryear{{Lidz}, {Zahn}, {Furlanetto}, {McQuinn},
  {Hernquist}  \& {Zaldarriaga}}{{Lidz} et~al.}{2009}]{2009ApJ...690..252L}
{Lidz} A.,  {Zahn} O.,  {Furlanetto} S.~R.,  {McQuinn} M.,  {Hernquist} L.,
  {Zaldarriaga} M.,  2009, \mn@doi [\apj] {10.1088/0004-637X/690/1/252}, \href
  {https://ui.adsabs.harvard.edu/abs/2009ApJ...690..252L} {690, 252}

\bibitem[\protect\citeauthoryear{Liu, Parsons  \& Trott}{Liu
  et~al.}{2014}]{PhysRevD.90.023019}
Liu A.,  Parsons A.~R.,   Trott C.~M.,  2014, \mn@doi [Phys. Rev. D]
  {10.1103/PhysRevD.90.023019}, 90, 023019

\bibitem[\protect\citeauthoryear{{Loeb} \& {Barkana}}{{Loeb} \&
  {Barkana}}{2001}]{2001ARA&A..39...19L}
{Loeb} A.,  {Barkana} R.,  2001, \mn@doi [\araa]
  {10.1146/annurev.astro.39.1.19}, \href
  {http://adsabs.harvard.edu/abs/2001ARA%26A..39...19L} {39, 19}

\bibitem[\protect\citeauthoryear{Loureiro et~al.,}{Loureiro
  et~al.}{2019}]{10.1093/mnras/stz191}
Loureiro A.,  et~al., 2019, \mn@doi [Monthly Notices of the Royal Astronomical
  Society] {10.1093/mnras/stz191}, 485, 326

\bibitem[\protect\citeauthoryear{{Madau} \& {Rees}}{{Madau} \&
  {Rees}}{2000}]{2000ApJ...542L..69M}
{Madau} P.,  {Rees} M.~J.,  2000, \mn@doi [\apj] {10.1086/312934}, \href
  {https://ui.adsabs.harvard.edu/#abs/2000ApJ...542L..69M} {542, L69}

\bibitem[\protect\citeauthoryear{{Majumdar}, {Mellema}, {Datta}, {Jensen},
  {Choudhury}, {Bharadwaj}  \& {Friedrich}}{{Majumdar}
  et~al.}{2014}]{2014MNRAS.443.2843M}
{Majumdar} S.,  {Mellema} G.,  {Datta} K.~K.,  {Jensen} H.,  {Choudhury} T.~R.,
   {Bharadwaj} S.,   {Friedrich} M.~M.,  2014, \mn@doi [\mnras]
  {10.1093/mnras/stu1342}, \href
  {http://adsabs.harvard.edu/abs/2014MNRAS.443.2843M} {443, 2843}

\bibitem[\protect\citeauthoryear{{Mason} et~al.,}{{Mason}
  et~al.}{2018}]{2018arXiv180101891M}
{Mason} C.~A.,  et~al., 2018, \mn@doi [\apjl] {10.3847/2041-8213/aabbab}, \href
  {http://adsabs.harvard.edu/abs/2018ApJ...857L..11M} {857, L11}

\bibitem[\protect\citeauthoryear{Matsakis \& Klock}{Matsakis \&
  Klock}{2014}]{Matsakis:2014:RL:2692956.2663188}
Matsakis N.~D.,  Klock II F.~S.,  2014, \mn@doi [Ada Lett.]
  {10.1145/2692956.2663188}, 34, 103

\bibitem[\protect\citeauthoryear{{McGreer}, {Mesinger}  \&
  {D'Odorico}}{{McGreer} et~al.}{2015}]{2015MNRAS.447..499M}
{McGreer} I.~D.,  {Mesinger} A.,   {D'Odorico} V.,  2015, \mn@doi [Monthly
  Notices of the Royal Astronomical Society] {10.1093/mnras/stu2449}, \href
  {https://ui.adsabs.harvard.edu/abs/2015MNRAS.447..499M} {447, 499}

\bibitem[\protect\citeauthoryear{{McQuinn}, {Zahn}, {Zaldarriaga}, {Hernquist}
  \& {Furlanetto}}{{McQuinn} et~al.}{2006}]{2006ApJ...653..815M}
{McQuinn} M.,  {Zahn} O.,  {Zaldarriaga} M.,  {Hernquist} L.,   {Furlanetto}
  S.~R.,  2006, \mn@doi [\apj] {10.1086/505167}, \href
  {https://ui.adsabs.harvard.edu/abs/2006ApJ...653..815M} {653, 815}

\bibitem[\protect\citeauthoryear{{McQuinn}, {Lidz}, {Zahn}, {Dutta},
  {Hernquist}  \& {Zaldarriaga}}{{McQuinn} et~al.}{2007a}]{2007MNRAS.377.1043M}
{McQuinn} M.,  {Lidz} A.,  {Zahn} O.,  {Dutta} S.,  {Hernquist} L.,
  {Zaldarriaga} M.,  2007a, \mn@doi [\mnras]
  {10.1111/j.1365-2966.2007.11489.x}, \href
  {https://ui.adsabs.harvard.edu/abs/2007MNRAS.377.1043M} {377, 1043}

\bibitem[\protect\citeauthoryear{{McQuinn}, {Hernquist}, {Zaldarriaga}  \&
  {Dutta}}{{McQuinn} et~al.}{2007b}]{2007MNRAS.381...75M}
{McQuinn} M.,  {Hernquist} L.,  {Zaldarriaga} M.,   {Dutta} S.,  2007b, \mn@doi
  [\mnras] {10.1111/j.1365-2966.2007.12085.x}, \href
  {https://ui.adsabs.harvard.edu/abs/2007MNRAS.381...75M} {381, 75}

\bibitem[\protect\citeauthoryear{Mesinger \& Furlanetto}{Mesinger \&
  Furlanetto}{2007}]{Mesinger:2007pd}
Mesinger A.,  Furlanetto S.,  2007, \mn@doi [Astrophys. J.] {10.1086/521806},
  669, 663

\bibitem[\protect\citeauthoryear{{Mesinger}, {Furlanetto}  \& {Cen}}{{Mesinger}
  et~al.}{2011}]{2011MNRAS.411..955M}
{Mesinger} A.,  {Furlanetto} S.,   {Cen} R.,  2011, \mn@doi [\mnras]
  {10.1111/j.1365-2966.2010.17731.x}, \href
  {http://adsabs.harvard.edu/abs/2011MNRAS.411..955M} {411, 955}

\bibitem[\protect\citeauthoryear{{Miralda-Escud{\'e}} \&
  {Rees}}{{Miralda-Escud{\'e}} \& {Rees}}{1994}]{1994MNRAS.266..343M}
{Miralda-Escud{\'e}} J.,  {Rees} M.~J.,  1994, \mn@doi [Monthly Notices of the
  Royal Astronomical Society] {10.1093/mnras/266.2.343}, \href
  {https://ui.adsabs.harvard.edu/abs/1994MNRAS.266..343M} {266, 343}

\bibitem[\protect\citeauthoryear{{Miralda-Escud{\'e}}, {Haehnelt}  \&
  {Rees}}{{Miralda-Escud{\'e}} et~al.}{2000}]{2000ApJ...530....1M}
{Miralda-Escud{\'e}} J.,  {Haehnelt} M.,   {Rees} M.~J.,  2000, \mn@doi [The
  Astrophysical Journal] {10.1086/308330}, \href
  {https://ui.adsabs.harvard.edu/abs/2000ApJ...530....1M} {530, 1}

\bibitem[\protect\citeauthoryear{{Miyazaki} et~al.,}{{Miyazaki}
  et~al.}{2018}]{2018PASJ...70S...1M}
{Miyazaki} S.,  et~al., 2018, \mn@doi [\pasj] {10.1093/pasj/psx063}, \href
  {https://ui.adsabs.harvard.edu/abs/2018PASJ...70S...1M} {70, S1}

\bibitem[\protect\citeauthoryear{{Morales}}{{Morales}}{2005}]{2005ApJ...619..678M}
{Morales} M.~F.,  2005, \mn@doi [\apj] {10.1086/426730}, \href
  {https://ui.adsabs.harvard.edu/abs/2005ApJ...619..678M} {619, 678}

\bibitem[\protect\citeauthoryear{{Moriwaki}, {Yoshida}, {Eide}  \&
  {Ciardi}}{{Moriwaki} et~al.}{2019}]{2019arXiv190610863M}
{Moriwaki} K.,  {Yoshida} N.,  {Eide} M.~B.,   {Ciardi} B.,  2019, arXiv
  e-prints, \href {https://ui.adsabs.harvard.edu/abs/2019arXiv190610863M} {p.
  arXiv:1906.10863}

\bibitem[\protect\citeauthoryear{Murray \& Poulin}{Murray \&
  Poulin}{2019}]{HANKEL}
Murray S.,  Poulin F.,  2019, \mn@doi [Journal of Open Source Software]
  {10.21105/joss.01397}, 4, 1397

\bibitem[\protect\citeauthoryear{{Nasir} \& {D'Aloisio}}{{Nasir} \&
  {D'Aloisio}}{2019}]{2019arXiv191003570N}
{Nasir} F.,  {D'Aloisio} A.,  2019, arXiv e-prints, \href
  {https://ui.adsabs.harvard.edu/abs/2019arXiv191003570N} {p. arXiv:1910.03570}

\bibitem[\protect\citeauthoryear{{Neben}, {Stalder}, {Hewitt}  \&
  {Tonry}}{{Neben} et~al.}{2017}]{2017ApJ...849...50N}
{Neben} A.~R.,  {Stalder} B.,  {Hewitt} J.~N.,   {Tonry} J.~L.,  2017, \mn@doi
  [\apj] {10.3847/1538-4357/aa8f9c}, \href
  {https://ui.adsabs.harvard.edu/abs/2017ApJ...849...50N} {849, 50}

\bibitem[\protect\citeauthoryear{{Ota} et~al.,}{{Ota}
  et~al.}{2017}]{2017arXiv170302501O}
{Ota} K.,  et~al., 2017, \mn@doi [\apj] {10.3847/1538-4357/aa7a0a}, \href
  {http://adsabs.harvard.edu/abs/2017ApJ...844...85O} {844, 85}

\bibitem[\protect\citeauthoryear{{Ouchi} et~al.,}{{Ouchi}
  et~al.}{2018}]{2017arXiv170407455O}
{Ouchi} M.,  et~al., 2018, \mn@doi [Publications of the Astronomical Society of
  Japan] {10.1093/pasj/psx074}, \href
  {https://ui.adsabs.harvard.edu/#abs/2018PASJ...70S..13O} {70, S13}

\bibitem[\protect\citeauthoryear{{Park}, {Kim}, {Wyithe}  \& {Lacey}}{{Park}
  et~al.}{2014}]{2014MNRAS.438.2474P}
{Park} J.,  {Kim} H.-S.,  {Wyithe} J. S.~B.,   {Lacey} C.~G.,  2014, \mn@doi
  [\mnras] {10.1093/mnras/stt2366}, \href
  {https://ui.adsabs.harvard.edu/abs/2014MNRAS.438.2474P} {438, 2474}

\bibitem[\protect\citeauthoryear{{Parsons}, {Pober}, {McQuinn}, {Jacobs}  \&
  {Aguirre}}{{Parsons} et~al.}{2012}]{2012ApJ...753...81P}
{Parsons} A.,  {Pober} J.,  {McQuinn} M.,  {Jacobs} D.,   {Aguirre} J.,  2012,
  \mn@doi [\apj] {10.1088/0004-637X/753/1/81}, \href
  {https://ui.adsabs.harvard.edu/abs/2012ApJ...753...81P} {753, 81}

\bibitem[\protect\citeauthoryear{{Planck Collaboration} et~al.,}{{Planck
  Collaboration} et~al.}{2014}]{2014A&A...571A..16P}
{Planck Collaboration} et~al., 2014, \mn@doi [\aap]
  {10.1051/0004-6361/201321591}, \href
  {http://adsabs.harvard.edu/abs/2014A%26A...571A..16P} {571, A16}

\bibitem[\protect\citeauthoryear{{Planck Collaboration}, {Aghanim}, {Akrami},
  {Ashdown}, {Aumont}  \& {et al.}}{{Planck Collaboration}
  et~al.}{2018}]{2018arXiv180706209P}
{Planck Collaboration} {Aghanim} N.,  {Akrami} Y.,  {Ashdown} M.,  {Aumont} J.,
    {et al.} 2018, preprint, \href
  {https://ui.adsabs.harvard.edu/#abs/2018arXiv180706209P} {p.
  arXiv:1807.06209} (\mn@eprint {arXiv} {1807.06209})

\bibitem[\protect\citeauthoryear{{Pritchard} \& {Loeb}}{{Pritchard} \&
  {Loeb}}{2012}]{2012RPPh...75h6901P}
{Pritchard} J.~R.,  {Loeb} A.,  2012, \mn@doi [Reports on Progress in Physics]
  {10.1088/0034-4885/75/8/086901}, \href
  {http://adsabs.harvard.edu/abs/2012RPPh...75h6901P} {75, 086901}

\bibitem[\protect\citeauthoryear{{Rahmati}, {Pawlik}, {Rai{\v c}evi{\`{c}}}  \&
  {Schaye}}{{Rahmati} et~al.}{2013}]{2013MNRAS.430.2427R}
{Rahmati} A.,  {Pawlik} A.~H.,  {Rai{\v c}evi{\`{c}}} M.,   {Schaye} J.,  2013,
  \mn@doi [\mnras] {10.1093/mnras/stt066}, \href
  {http://adsabs.harvard.edu/abs/2013MNRAS.430.2427R} {430, 2427}

\bibitem[\protect\citeauthoryear{{Rosdahl} et~al.,}{{Rosdahl}
  et~al.}{2018}]{2018MNRAS.479..994R}
{Rosdahl} J.,  et~al., 2018, \mn@doi [\mnras] {10.1093/mnras/sty1655}, \href
  {https://ui.adsabs.harvard.edu/#abs/2018MNRAS.479..994R} {479, 994}

\bibitem[\protect\citeauthoryear{{SKAO Science Team}}{{SKAO Science
  Team}}{2015}]{SKA-SCI-LOW-001}
{SKAO Science Team} 2015, Technical Report SKA-SCI-LOW-001, SKA1-Low
  Configuration.
SKA Organisation

\bibitem[\protect\citeauthoryear{{Sadoun}, {Zheng}  \&
  {Miralda-Escud{\'e}}}{{Sadoun} et~al.}{2017}]{2017ApJ...839...44S}
{Sadoun} R.,  {Zheng} Z.,   {Miralda-Escud{\'e}} J.,  2017, \mn@doi [\apj]
  {10.3847/1538-4357/aa683b}, \href
  {http://adsabs.harvard.edu/abs/2017ApJ...839...44S} {839, 44}

\bibitem[\protect\citeauthoryear{{Santos}, {Ferramacho}, {Silva}, {Amblard}  \&
  {Cooray}}{{Santos} et~al.}{2010}]{2010MNRAS.406.2421S}
{Santos} M.~G.,  {Ferramacho} L.,  {Silva} M.~B.,  {Amblard} A.,   {Cooray} A.,
   2010, \mn@doi [\mnras] {10.1111/j.1365-2966.2010.16898.x}, \href
  {http://adsabs.harvard.edu/abs/2010MNRAS.406.2421S} {406, 2421}

\bibitem[\protect\citeauthoryear{{Shibuya} et~al.,}{{Shibuya}
  et~al.}{2018}]{2018PASJ...70S..14S}
{Shibuya} T.,  et~al., 2018, \mn@doi [Publications of the Astronomical Society
  of Japan] {10.1093/pasj/psx122}, \href
  {https://ui.adsabs.harvard.edu/#abs/2018PASJ...70S..14S} {70, S14}

\bibitem[\protect\citeauthoryear{{Shimabukuro} \& {Semelin}}{{Shimabukuro} \&
  {Semelin}}{2017}]{2017MNRAS.468.3869S}
{Shimabukuro} H.,  {Semelin} B.,  2017, \mn@doi [\mnras]
  {10.1093/mnras/stx734}, \href
  {http://adsabs.harvard.edu/abs/2017MNRAS.468.3869S} {468, 3869}

\bibitem[\protect\citeauthoryear{{Sims}, {Lentati}, {Alexander}  \&
  {Carilli}}{{Sims} et~al.}{2016}]{2016MNRAS.462.3069S}
{Sims} P.~H.,  {Lentati} L.,  {Alexander} P.,   {Carilli} C.~L.,  2016, \mn@doi
  [\mnras] {10.1093/mnras/stw1768}, \href
  {https://ui.adsabs.harvard.edu/abs/2016MNRAS.462.3069S} {462, 3069}

\bibitem[\protect\citeauthoryear{{Smith}, {Sheth}  \& {Scoccimarro}}{{Smith}
  et~al.}{2008}]{2008PhRvD..78b3523S}
{Smith} R.~E.,  {Sheth} R.~K.,   {Scoccimarro} R.,  2008, \mn@doi [\prd]
  {10.1103/PhysRevD.78.023523}, \href
  {https://ui.adsabs.harvard.edu/abs/2008PhRvD..78b3523S} {78, 023523}

\bibitem[\protect\citeauthoryear{{Sobacchi}, {Mesinger}  \& {Greig}}{{Sobacchi}
  et~al.}{2016}]{2016MNRAS.459.2741S}
{Sobacchi} E.,  {Mesinger} A.,   {Greig} B.,  2016, \mn@doi [\mnras]
  {10.1093/mnras/stw811}, \href
  {https://ui.adsabs.harvard.edu/abs/2016MNRAS.459.2741S} {459, 2741}

\bibitem[\protect\citeauthoryear{{Spergel} et~al.,}{{Spergel}
  et~al.}{2015}]{2015arXiv150303757S}
{Spergel} D.,  et~al., 2015, arXiv e-prints, \href
  {https://ui.adsabs.harvard.edu/abs/2015arXiv150303757S} {p. arXiv:1503.03757}

\bibitem[\protect\citeauthoryear{Springel}{Springel}{2005}]{GADGET-2}
Springel V.,  2005, \mn@doi [Monthly Notices of the Royal Astronomical Society]
  {10.1111/j.1365-2966.2005.09655.x}, 364, 1105

\bibitem[\protect\citeauthoryear{{Springel}, {Yoshida}  \& {White}}{{Springel}
  et~al.}{2001}]{2001NewA....6...79S}
{Springel} V.,  {Yoshida} N.,   {White} S.~D.~M.,  2001, \mn@doi [\na]
  {10.1016/S1384-1076(01)00042-2}, \href
  {http://adsabs.harvard.edu/abs/2001NewA....6...79S} {6, 79}

\bibitem[\protect\citeauthoryear{{Takada} et~al.,}{{Takada}
  et~al.}{2014}]{2014PASJ...66R...1T}
{Takada} M.,  et~al., 2014, \mn@doi [\pasj] {10.1093/pasj/pst019}, \href
  {https://ui.adsabs.harvard.edu/abs/2014PASJ...66R...1T} {66, R1}

\bibitem[\protect\citeauthoryear{{Tamura} et~al.,}{{Tamura}
  et~al.}{2018}]{2018SPIE10702E..1CT}
{Tamura} N.,  et~al., 2018, in Ground-based and Airborne Instrumentation for
  Astronomy VII. p. 107021C, \mn@doi{10.1117/12.2311871}

\bibitem[\protect\citeauthoryear{{Tegmark}, {Silk}, {Rees}, {Blanchard}, {Abel}
   \& {Palla}}{{Tegmark} et~al.}{1997}]{1997ApJ...474....1T}
{Tegmark} M.,  {Silk} J.,  {Rees} M.~J.,  {Blanchard} A.,  {Abel} T.,   {Palla}
  F.,  1997, \mn@doi [\apj] {10.1086/303434}, \href
  {https://ui.adsabs.harvard.edu/abs/1997ApJ...474....1T} {474, 1}

\bibitem[\protect\citeauthoryear{{Tingay} et~al.,}{{Tingay}
  et~al.}{2013}]{2013PASA...30....7T}
{Tingay} S.~J.,  et~al., 2013, \mn@doi [\pasa] {10.1017/pasa.2012.007}, \href
  {http://adsabs.harvard.edu/abs/2013PASA...30....7T} {30, e007}

\bibitem[\protect\citeauthoryear{{Tozzi}, {Madau}, {Meiksin}  \&
  {Rees}}{{Tozzi} et~al.}{2000}]{2000ApJ...528..597T}
{Tozzi} P.,  {Madau} P.,  {Meiksin} A.,   {Rees} M.~J.,  2000, \mn@doi [\apj]
  {10.1086/308196}, \href
  {https://ui.adsabs.harvard.edu/abs/2000ApJ...528..597T} {528, 597}

\bibitem[\protect\citeauthoryear{{Trac} \& {Cen}}{{Trac} \&
  {Cen}}{2007}]{2007ApJ...671....1T}
{Trac} H.,  {Cen} R.,  2007, \mn@doi [\apj] {10.1086/522566}, \href
  {https://ui.adsabs.harvard.edu/abs/2007ApJ...671....1T} {671, 1}

\bibitem[\protect\citeauthoryear{{Trenti}, {Stiavelli}, {Bouwens}, {Oesch},
  {Shull}, {Illingworth}, {Bradley}  \& {Carollo}}{{Trenti}
  et~al.}{2010}]{2010ApJ...714L.202T}
{Trenti} M.,  {Stiavelli} M.,  {Bouwens} R.~J.,  {Oesch} P.,  {Shull} J.~M.,
  {Illingworth} G.~D.,  {Bradley} L.~D.,   {Carollo} C.~M.,  2010, \mn@doi
  [\apjl] {10.1088/2041-8205/714/2/L202}, \href
  {http://adsabs.harvard.edu/abs/2010ApJ...714L.202T} {714, L202}

\bibitem[\protect\citeauthoryear{Van~Rossum \& Drake~Jr}{Van~Rossum \&
  Drake~Jr}{1995}]{van1995python}
Van~Rossum G.,  Drake~Jr F.~L.,  1995, Python tutorial.
Centrum voor Wiskunde en Informatica Amsterdam, The Netherlands

\bibitem[\protect\citeauthoryear{{Vrbanec} et~al.,}{{Vrbanec}
  et~al.}{2016}]{2016MNRAS.457..666V}
{Vrbanec} D.,  et~al., 2016, \mn@doi [\mnras] {10.1093/mnras/stv2993}, \href
  {https://ui.adsabs.harvard.edu/abs/2016MNRAS.457..666V} {457, 666}

\bibitem[\protect\citeauthoryear{{Waterson}, {Labate}, {Schnetler}, {Wagg},
  {Turner}  \& {Dewdney}}{{Waterson} et~al.}{2016}]{2016SPIE.9906E..28W}
{Waterson} M.~F.,  {Labate} M.~G.,  {Schnetler} H.,  {Wagg} J.,  {Turner} W.,
  {Dewdney} P.,  2016, in \procspie. p. 990628, \mn@doi{10.1117/12.2232526}

\bibitem[\protect\citeauthoryear{{Wayth} et~al.,}{{Wayth}
  et~al.}{2018}]{2018PASA...35...33W}
{Wayth} R.~B.,  et~al., 2018, \mn@doi [\pasa] {10.1017/pasa.2018.37}, \href
  {https://ui.adsabs.harvard.edu/abs/2018PASA...35...33W} {35, 33}

\bibitem[\protect\citeauthoryear{{Weinberger}, {Kulkarni}, {Haehnelt},
  {Choudhury}  \& {Puchwein}}{{Weinberger} et~al.}{2018}]{2018MNRAS.tmp.1485W}
{Weinberger} L.~H.,  {Kulkarni} G.,  {Haehnelt} M.~G.,  {Choudhury} T.~R.,
  {Puchwein} E.,  2018, \mn@doi [\mnras] {10.1093/mnras/sty1563}, \href
  {https://ui.adsabs.harvard.edu/#abs/2018MNRAS.479.2564W} {479, 2564}

\bibitem[\protect\citeauthoryear{{Weinberger}, {Haehnelt}  \&
  {Kulkarni}}{{Weinberger} et~al.}{2019}]{2019MNRAS.485.1350W}
{Weinberger} L.~H.,  {Haehnelt} M.~G.,   {Kulkarni} G.,  2019, \mn@doi [\mnras]
  {10.1093/mnras/stz481}, \href
  {https://ui.adsabs.harvard.edu/abs/2019MNRAS.485.1350W} {485, 1350}

\bibitem[\protect\citeauthoryear{{Wiersma} et~al.,}{{Wiersma}
  et~al.}{2013}]{2013MNRAS.432.2615W}
{Wiersma} R.~P.~C.,  et~al., 2013, \mn@doi [\mnras] {10.1093/mnras/stt624},
  \href {https://ui.adsabs.harvard.edu/abs/2013MNRAS.432.2615W} {432, 2615}

\bibitem[\protect\citeauthoryear{{Wild}}{{Wild}}{1952}]{1952ApJ...115..206W}
{Wild} J.~P.,  1952, \mn@doi [\apj] {10.1086/145533}, \href
  {https://ui.adsabs.harvard.edu/abs/1952ApJ...115..206W} {115, 206}

\bibitem[\protect\citeauthoryear{{Witstok}, {Puchwein}, {Kulkarni}, {Smit}  \&
  {Haehnelt}}{{Witstok} et~al.}{2019}]{2019arXiv190506954W}
{Witstok} J.,  {Puchwein} E.,  {Kulkarni} G.,  {Smit} R.,   {Haehnelt} M.~G.,
  2019, arXiv e-prints, \href
  {https://ui.adsabs.harvard.edu/abs/2019arXiv190506954W} {p. arXiv:1905.06954}

\bibitem[\protect\citeauthoryear{{Wyithe} \& {Morales}}{{Wyithe} \&
  {Morales}}{2007}]{2007MNRAS.379.1647W}
{Wyithe} J. S.~B.,  {Morales} M.~F.,  2007, \mn@doi [\mnras]
  {10.1111/j.1365-2966.2007.12048.x}, \href
  {https://ui.adsabs.harvard.edu/abs/2007MNRAS.379.1647W} {379, 1647}

\bibitem[\protect\citeauthoryear{{Yoshiura}, {Line}, {Kubota}, {Hasegawa}  \&
  {Takahashi}}{{Yoshiura} et~al.}{2018}]{2018MNRAS.479.2767Y}
{Yoshiura} S.,  {Line} J.~L.~B.,  {Kubota} K.,  {Hasegawa} K.,   {Takahashi}
  K.,  2018, \mn@doi [\mnras] {10.1093/mnras/sty1472}, \href
  {https://ui.adsabs.harvard.edu/abs/2018MNRAS.479.2767Y} {479, 2767}

\bibitem[\protect\citeauthoryear{{Zahn}, {Lidz}, {McQuinn}, {Dutta},
  {Hernquist}, {Zaldarriaga}  \& {Furlanetto}}{{Zahn}
  et~al.}{2007}]{2007ApJ...654...12Z}
{Zahn} O.,  {Lidz} A.,  {McQuinn} M.,  {Dutta} S.,  {Hernquist} L.,
  {Zaldarriaga} M.,   {Furlanetto} S.~R.,  2007, \mn@doi [\apj]
  {10.1086/509597}, \href {http://adsabs.harvard.edu/abs/2007ApJ...654...12Z}
  {654, 12}

\bibitem[\protect\citeauthoryear{{Zheng} \& {Wallace}}{{Zheng} \&
  {Wallace}}{2014}]{2014ApJ...794..116Z}
{Zheng} Z.,  {Wallace} J.,  2014, \mn@doi [\apj] {10.1088/0004-637X/794/2/116},
  \href {http://adsabs.harvard.edu/abs/2014ApJ...794..116Z} {794, 116}

\bibitem[\protect\citeauthoryear{{van Haarlem} et~al.,}{{van Haarlem}
  et~al.}{2013}]{2013A&A...556A...2V}
{van Haarlem} M.~P.,  et~al., 2013, \mn@doi [\aap]
  {10.1051/0004-6361/201220873}, \href
  {http://adsabs.harvard.edu/abs/2013A%26A...556A...2V} {556, A2}

\bibitem[\protect\citeauthoryear{van~der Walt, Colbert  \& Varoquaux}{van~der
  Walt et~al.}{2011}]{NUMPY}
van~der Walt S.,  Colbert S.~C.,   Varoquaux G.,  2011, \mn@doi [Computing in
  Science Engineering] {10.1109/MCSE.2011.37}, 13, 22

\makeatother
\end{thebibliography}



\appendix
\section{Power spectrum estimation}
\label{appendix:ps_details}
We follow the approach of \citet{2008PhRvD..78b3523S} to estimate the power spectrum\footnote{A Rust implementation of this calculation is publicly available at \url{https://github.com/lewis-weinberger/crosscorr}.}. Computationally we calculate the Fourier transforms of our simulation quantities using Fast Fourier Transforms (FFT), taking advantage of the \texttt{FFTW} library \citep{FFTW05}. We correct for the grid discreteness by dividing out the Fourier transform of the grid interpolation window function,
\begin{equation}
\tilde{\delta} = \tilde{\delta}(\mathbf{k}) / W_{\rm CIC} (\mathbf{k}),
\end{equation}
where $W_{\rm CIC} (\mathbf{k})$ is the Fourier transform of the Cloud-In-Cell (CIC) window function,
\begin{multline}
W_{\rm CIC}(\mathbf{k}) = \\
\left[ \sinc\left(\frac{\pi k_1}{2 k_{\rm Ny}}\right)
\sinc\left(\frac{\pi k_2}{2 k_{\rm Ny}}\right)
\sinc\left(\frac{\pi k_3}{2 k_{\rm Ny}}\right)\right]^2,
\end{multline}
where $k_{\rm Ny} = \pi N_{\rm grid}/ L_{\rm box}$ is the Nyquist frequency (for $N_{\rm grid}$ cells on a side of length $L_{\rm box}$).

At a scale $k$ the spherically-averaged auto-power spectrum can be estimated as,
\begin{align}
  \hat{P}(k) &= \frac{V}{N_m} \sum_{i=1}^{N_m} |\tilde{\delta}(\mathbf{k}_i)|^2, \\
             &=  \frac{V}{N_{\rm grid}^6} \left(\frac{1}{N_m} \sum_{i=1}^{N_m} |\tilde{\delta}^{\tt FFTW}(\mathbf{k}_i)|^2 \right),
\end{align}
where $N_m$ is the number of modes in a spherical shell and $V$ is the simulation volume, $V = L_{\rm box}^3$. Note here that the \texttt{FFTW} transform is an unnormalised discrete Fourier transform (DFT), hence we must include the appropriate normalisation $\tilde{\delta} = \tilde{\delta}^{\tt FFTW}/N_{\rm grid}^3$. Similarly for the cross-power spectrum,
\begin{equation}
\hat{P}_{a,b}(k) = \frac{V}{N_m} \sum_{i=1}^{N_m} \frac{[\tilde{\delta}_a^*(\mathbf{k}_i) \tilde{\delta}_b(\mathbf{k}_i) + \tilde{\delta}_a(\mathbf{k}_i) \tilde{\delta}_b^*(\mathbf{k}_i)]}{2},
\end{equation}
where $\tilde{\delta}^*$ indicates the complex conjugate.

We correct for shot noise when calculating power spectrum, $P_{\rm LAE} (k)$,
following \citet{2008PhRvD..78b3523S},
\begin{equation}
\hat{P}(k) = \hat{P}(k) - \hat{P}^{\rm shot}(k).
\end{equation}
For the LAE auto-power spectrum the shot noise term is given by,
\begin{equation}
\hat{P}_{\rm LAE \times LAE}^{\rm shot}(k) = \frac{V}{N_{\rm obj}},
\end{equation}
where $N_{\rm obj}$ is the number of objects (e.g. $N_{\rm LAE}$) in the simulation volume. Note the 21cm-LAE cross-power spectrum estimator does not need to be corrected for LAE shot noise.
The shot noise corrections affect small scales, $k > 1$ h/cMpc, and are strongly dependent on the number of observed LAEs, $N_{\rm LAE}$.

\section{Survey sensitivities}
\label{appendix:surveys}
In this appendix we derive expressions for the noise terms used in Eqs.~(\ref{eq:21cmtot}) \&~(\ref{eq:laetot}).

\subsection{21cm surveys}
\label{appendix:21cmsurveys}
We follow the treatments of \citet{2012ApJ...753...81P}, \citet{2011MNRAS.418..516G} and \citet{2006ApJ...653..815M} in order to derive the thermal noise of 21-cm observations. We neglect any systematic errors introduced by foreground removal \citep[and refer readers to][for further exploration of foreground effects]{2018MNRAS.479.2767Y}. The variance in a measurement of the 21-cm power spectrum is given by,
\begin{equation}
    {\rm var}[P_{\rm 21cm}(k,\mu)] = \left[P_{\rm 21cm}(k,\mu) + P^{\rm noise}_{\rm 21cm}(k,\mu) \right]^2,
\end{equation}
 where the first term on the right-hand side is the contribution from sample variance (i.e. random error), and the second term is the contribution from thermal noise in the interferometer (i.e. systematic error).

\subsubsection{Deriving the power spectrum from interferometric visibilities}
Interferometers measure fringe visibilities, $V(\mathbf{U}, \nu)$, where $\mathbf{U} = (u, v)$ is the image coordinate measured in wavelengths (sometimes called the \textit{spatial frequency}), and $\nu$ is the frequency of the measurement. Note we are assuming the flat-sky approximation in which we neglect line-of-sight dependence. We will use the notation $\tilde{V}(u,v,\eta)$ to refer to the frequency Fourier transform of the visibility ($\eta$ is an inverse frequency). The observed power spectrum of brightness temperature fluctuations can be approximately related to $\tilde{V}$ by \citep[see][for a detailed derivation]{2012ApJ...753...81P},
\begin{multline}
\label{eq:parsons}
P(k,\mu) = \langle|\tilde{T}_b(\mathbf{k})|^2\rangle \simeq \\
\left[\left(\frac{\lambda^2}{2 k_B}\right)^2 \frac{D^2\: \Delta D}{\Omega \: B^2}\right]\: \tilde{V}^2(u,v,\eta)
\end{multline}
where $\lambda$ is the observed wavelength, $k_B$ is Boltzmann's constant, $D$ is the comoving distance to the surveyed emission, $\Delta D$ is the comoving radial depth of the survey, $B$ is the bandwidth of the measurement, and $\Omega$ is the field of view of the telescope given by $\Omega = \lambda^2/A_e$, where $A_e$ is the effective area of an antenna. The Fourier mode $(u, v, \eta)$ measured by the inteferometer is related to the cosmological $\mathbf{k}$-mode by,
\begin{equation}
(Dk_x, Dk_y, \Delta D k_z/B) = 2\pi(u, v, \eta),
\end{equation}
or equivalently,
\begin{equation}
  \label{eq:mapping}
Dk_{\bot} = 2\pi U, \quad \Delta Dk_{\parallel}/B = 2\pi\eta,
\end{equation}
where we have split the $\mathbf{k}$-mode into components parallel and perpendicular to the line-of-sight of the observation, $\mathbf{k} = \mathbf{k}_{\bot} + \mathbf{k}_{\parallel}$, and $U = |\mathbf{U}|$.

\subsubsection{Thermal detector noise}
We wish to derive the thermal noise of a given k-mode that is left \emph{after} attempting to subtract the noise power (using any available information). In time $t_\mathbf{k}$ we can make $2Bt_\mathbf{k}$ independent measurements of the \emph{observation-independent} system temperature, $T_{\rm sys}$. Assuming Gaussian random fluctuations in our measurements of the system temperature (which have magnitude $\sqrt{2}T_{\rm sys}$), the thermal fluctuations in the detector have \emph{observation-dependent} rms brightness temperature,
\begin{equation}
{(T^{\rm noise}_{\rm rms})}^2 = \frac{(\sqrt{2}T_{\rm sys})^2}{2Bt_\mathbf{k}}
= \frac{T^2_{\rm sys}}{B t_\mathbf{k}},
\end{equation}
which is also known as the \emph{radiometer equation}.
This adds a white-noise contribution to the rms amplitude of the Fourier transform of the visibility given by,
\begin{equation}
\tilde{V}^{\rm noise}(u, v, \eta) = \frac{2 k_B}{\lambda^2} \Omega \: B \: {T^{\rm noise}_{\rm rms}}(u, v, \eta)
\end{equation}
Using Eq.~(\ref{eq:parsons}) the noise power is therefore given by \citep{2012ApJ...753...81P,2006ApJ...653..815M,2005ApJ...619..678M},
\begin{equation}
\label{eq:ptot_21}
P^{\rm noise}_{\rm 21cm}(k, \mu) =
D^2 \Delta D \frac{\Omega\: T^2_{\rm sys}}{2 \:B\: t_{\mathbf{k}}},
\end{equation}
where we have introduced a factor of two in the denominator to account for using two orthogonal polarization measurements to measure the total signal.

We note that $t_\mathbf{k}$ is the average time that a given \textbf{k}-mode can be observed by the interferometer (not simply the total integration time), which will depend explicitly on the configuration and position (latitude) of a given array.  Here we estimate this time as \citep{2006ApJ...653..815M},
\begin{equation}
\label{eq:tk}
t_\mathbf{k} \simeq \:t_{\rm int}\: n_b(k_{\bot}) / \Omega,
\end{equation}
where $t_{\rm int}$ is the total integration time of the observation and $n_b(k_{\bot})$ is the number density of baselines observing at a given time that sample a particular transverse k-mode, $k_{\bot} = k\sqrt{1 - \mu^2}$, which accounts for the antenna array geometry.

The final expression for the thermal noise on a $(k, \mu)$-mode is therefore given by,
\begin{equation}
\label{eq:thermal_noise}
P^{\rm noise}_{\rm 21cm}(k, \mu) =
\frac{D^2 \Delta D}{n_b(k_\bot)}\frac{\Omega^2\: T^2_{\rm sys}}{2 \:B\: t_{\rm int}}.
\end{equation}
We assume a system temperature dominated by the sky temperature \citep{2007MNRAS.379.1647W},
\begin{equation}
T_{\rm sys} \simeq 280 \mathrm{K} \left(\frac{1 + z}{7.5}\right)^{2.3}.
\end{equation}
We note that the comoving survey depth is approximately given by \citep{2012ApJ...753...81P},
\begin{equation}
\Delta D \simeq 1.7 \left(\frac{B}{0.1 \rm\: MHz} \right) \left(\frac{1 + z}{10}\right)^{0.5}
\left(\frac{\Omega_{\rm m} h^2}{0.15}\right)^{-1.5} \: \textrm{cMpc}.
\end{equation}
For all telescopes we assume an integration time of $t_{\rm int} = 1000$ hrs, a bandpass of $B=8$ MHz (which corresponds to depth of $\Delta D(z=7) \simeq 100$ cMpc), and a frequency resolution of $\Delta\nu=50$ kHz. And finally we note that using the above quantities the 21-cm survey volume is given by,
\begin{equation}
  V_{\rm survey} = D^2 \Delta D \:\Omega = D^2 \Delta D \left(\frac{\lambda^2}{A_e}\right)
  \label{eq:21cm_vol}
\end{equation}
We list the 21-cm survey parameters that we use to estimate sensitivities in Table~\ref{tab:surveys}.

\subsubsection{Baseline number density}
The baseline number density, $n_b(k_\bot)$, is equivalently parameterised by $U$ (following Eq.~(\ref{eq:mapping})), such that we can also write it as $n_b(U)$. This density is normalised so that the integral of $n_b(U){\rm d}U$ over the observed half-plane is the total number of baselines,
\begin{equation}
\label{eq:norm}
N_{\rm baselines} = \frac{1}{2} N_a ( N_a - 1),
\end{equation}
where $N_a$ is the total number of antennae\footnote{In practice each ``antenna tile'' is a collection of individual antenna elements, such as a phased-array of dipole antennae. This collection is often referred to simply as an antenna, or a station.} in the array.
If we approximate that the array is circularly symmetric, we can calculate the number density of baselines as a convolution \citep{2011MNRAS.418..516G,2007MNRAS.382..809D},
\begin{multline}
  \label{eq:convolution}
  n_b(U) = \\ C_b \int^{r_{\rm max}}_{0}  \mathrm{d}r \: 2 \pi r \: n_a(r)
  \int^{2 \pi}_{0} \mathrm{d}\phi\: n_a(|\mathbf{r} - \lambda \mathbf{U}|),
\end{multline}
where $n_a$ is the number density of antennae, and $C_b$ is a (frequency-dependent) normalisation constant that ensures the integral over the half-plane is given by Eq.~(\ref{eq:norm}),
\begin{equation}
 N_{\rm baselines} = \int_{0}^{U_{\rm max}} {\rm d}U\: 2\pi \:U \:n_b(U)
\end{equation}
Using Eq.~(\ref{eq:mapping}) we can rewrite,
\begin{equation}
|\mathbf{r} - \lambda \mathbf{U}| = \sqrt{r^2 - 2 \chi r k_{\bot} \cos{\phi}  + \chi^2 k_{\bot}^2},
\end{equation}
where $\chi = \lambda U/k_{\bot} = \lambda D/ 2\pi$.

We can perform the convolution in Eq.~(\ref{eq:convolution}) numerically using the actual array configuration, $n_a(r)$, of a given interferometer, or alternatively using an idealised model. In this work we use an idealised model following the configuration described in \citet{2011MNRAS.418..516G},
\begin{equation}
  n_a(r) =
  \begin{cases}
    n_{\rm core} & r \leq r_{\rm core}, \\
    n_{\rm core}\left(r_{\rm core}/r\right)^2 & r_{\rm core} < r < r_{\rm max}, \\
    0 & {\rm otherwise},
  \end{cases}
\end{equation}
where $n_{\rm core}$ is the number density of a core region within radius $r_{\rm core}$, given by,
\begin{equation}
  n_{\rm core} = \frac{N_a}{\pi r_{\rm core}^2 [ 1 + 2 \ln{(r_{\rm max}/r_{\rm core})}]}.
\end{equation}
The number density in the core region is limited by the physical size of the antenna tiles, such that $n_{\rm core} < {1}/{A_e}$. We choose values for the parameters $N_a$, $r_{\rm core}$ and $r_{\rm max}$ that reflect the configurations of MWA-Phase II (extended configuration), LOFAR, HERA and SKA1-low, similar to those of \citet{2016MNRAS.463.2583K}. These values can be found in Table~\ref{tab:surveys}.

\subsection{LAE surveys}
We follow the treatments of \citet{1994ApJ...426...23F} and  \citet{1997ApJ...474....1T} in order to derive errors on the LAE galaxy survey. As for the 21-cm case, the variance in a measurement of the LAE power spectrum is given by,
\begin{equation}
  \label{eq:laevar1}
    {\rm var}[P_{\rm LAE}(k,\mu)] = \left[P_{\rm LAE}(k,\mu) + P^{\rm noise}_{\rm LAE}(k,\mu) \right]^2,
\end{equation}
where as before the first term on the right-hand side is the contribution from sample variance, and the second term is the contribution from galaxy shot noise.

\subsubsection{Shot noise}
If we include the Poissonian shot noise term explicitly, Eq.~(\ref{eq:laevar1}) can be rewritten \citep{1994ApJ...426...23F} (dropping the $k,\mu$ labels),
\begin{equation}
  \label{eq:laevar}
  \frac{{\rm var}[P_{\rm LAE}]}{P_{\rm LAE}} =
  \left[ 1 + \frac{1}{\bar{n}_{\rm LAE}\: P_{\rm LAE}} \right]^2,
\end{equation}
where $\bar{n}_{\rm LAE}$ is the mean LAE number density in the observed sample.

\subsubsection{Redshift measurements}
We assume the redshift estimates (photometric from HSC or spectroscopic from PFS) introduce a Gaussian spread in the redshift-space measurement of $n_{\rm LAE}$ along the line-of-sight \citep{10.1093/mnras/stz191},
\begin{equation}
n_{\rm LAE}(z) = \int {\rm d}z' n^{\rm int}_{\rm LAE}(z - z') \exp{\left(-\frac{{z'}^2}{2 \sigma_z^2} \right)},
\end{equation}
where $n^{\rm int}_{\rm LAE}(z)$ is the underlying, intrinsic redshift distribution of the LAEs, and $\sigma_z$ is the typical redshift measurement error. This distorts the measured power spectrum (estimated in redshift space),
\begin{equation}
  P(k, \mu) \rightarrow P(k,\mu) \:\exp{\left(-\left[k_{\parallel} c \sigma_z / H(z)\right]^2\right)},
\end{equation}
where $k_{\parallel} = \mu k$ is the component of the wave-vector parallel to the line-of-sight, $c$ is the speed of light, and $H(z)$ is the Hubble parameter. Including this effect in Eq.~(\ref{eq:laevar}), we find that the shot noise term is increased by a factor $\exp{(\left[k_{\parallel} c \sigma_z / H(z)\right]^2)}$.

\subsubsection{Combined error}
The final shot noise error term for a $(k,\mu)$-mode of the LAE galaxy survey is therefore given by,
\begin{equation}
\label{eq:ptot_lae}
P^{\rm noise}_{\rm LAE} (k, \mu) = \frac{\exp{(\left[k_{\parallel} c \sigma_z / H(z)\right]^2)}}{\bar{n}_{\rm LAE}}.
\end{equation}
As outlined in section~\ref{sec:method_obs}, we estimate the sensitivity of a survey similar to SILVERRUSH \citep{2017arXiv170407455O}, performed using the Subaru telescope. We can estimate $\bar{n}_{\rm LAE}$ for a given depth by integrating the luminosity functions observed in the SILVERRUSH survey; in practice we use the Schechter fits from \citet{2018arXiv180505944I}. We choose $\sigma_z = 0.0007$ for the PFS redshift errors \citep{2014PASJ...66R...1T}, following \citet{2018MNRAS.479.2754K}.

\subsection{Volume matching and available modes}
\label{appendix:volume}
In practice a cross-correlation study between 21-cm and LAE surveys would require careful volume matching to ensure that the same physical region of the Universe is being sampled by each telescope. In order to replicate this in our sensitivity estimates we limit the included modes of the surveys to that of the more restrictive survey, and also set $V_{\rm survey}$ in Eq.~(\ref{eq:nmodes}) to be the smaller of the two volumes $\{V_{\rm 21cm}, V_{\rm LAE}\}$. For the telescopes considered in this work, and for a given $\mathbf{k}$-mode, we note that the $\mu$-modes available for extraction are limited by the 21-cm surveys \citep{2007ApJ...660.1030F}. In particular there is a maximum parallel component set by the depth of the survey,
\begin{equation}
\mu_{\rm max} = \mathrm{min}(1, k/k_{\rm \parallel, min}),
\end{equation}
and a minimum transverse component given by,
\begin{equation}
\mu_{\rm min}^2 = \mathrm{max}(0, 1 - k_{\rm \bot, max}^2/k^2),
\end{equation}
where $k_{\rm \parallel, min}=2\pi/\Delta D$ and $k_{\rm \bot, max}=2\pi u_{\rm max}/D$, for maximum baseline spatial frequency $u_{\rm max} = |\mathbf{u}|_{\rm max}$.

In Table~\ref{tab:surveys} we list the survey volumes and available modes for each of the surveys considered in this work at $z=6.6$. The LAE survey volumes are calculated assuming a wide-field narrowband geometry,
\begin{equation}
  V_{\rm LAE} \simeq d(z)^3 A_{\theta} \left( \frac{d(z + \Delta z)}{d(z)} - 1 \right),
\end{equation}
where $d(z)$ is the comoving distance to a redshift $z$, the narrowband depth is assumed to be $\Delta z = 0.1$, and the angular area of the survey is $A_{\theta} = 27$ deg$^2$ (Subaru Deep field). The 21cm volumes are calculated using Eq.~(\ref{eq:21cm_vol}).

\section{Correlation functions}
  \label{appendix:C}
  \begin{figure*}
  \includegraphics[width=\textwidth]{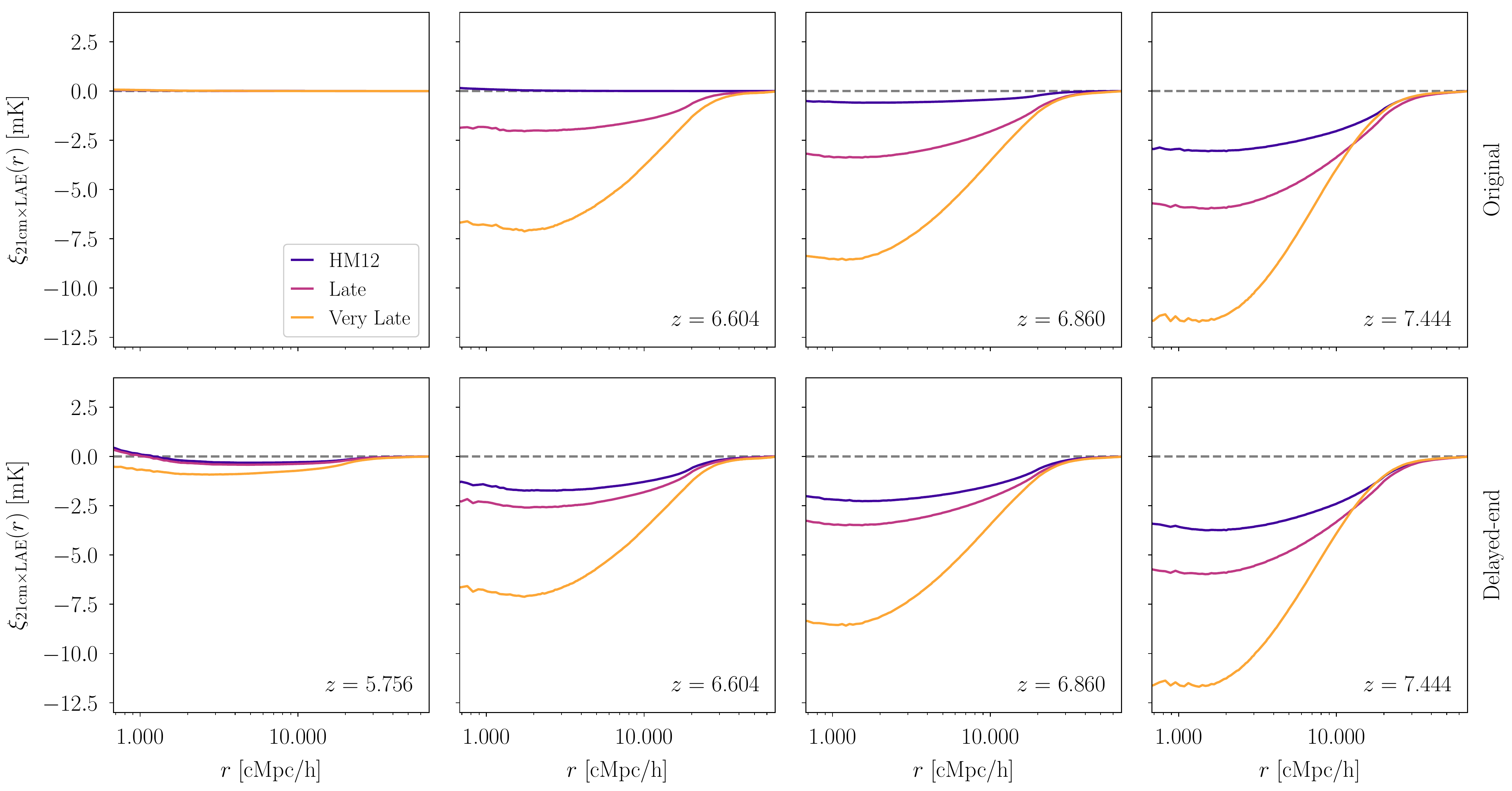}
  \caption{Cross-correlation functions for the reionization histories: HM12 (blue), Late (purple), Very Late (orange). The top panels show the original reionization histories, whilst the bottom panels show the delayed-end versions. From left to right the panels show increasing redshift, starting at $z=5.756$ on the left and ending with $z=7.444$ on the right. The dashed horizontal line indicates $\xi(r) =0$.}
  \label{fig:6}
\end{figure*}

  \subsection{Cross-correlation functions}
  We can use the $k$-space cross-power spectra to derive the real-space correlation functions by taking the Fourier transform,
\begin{equation}
\xi (r) = \frac{1}{(2 \pi)^3} \int P(k)\sinc(k r) 4 \pi k^2 \mathrm{d}k,
\end{equation}
where $\sinc(x) = \sin(x)/x$. We note that the integrand in this transform is highly oscillatory, and so the numerical integration must be performed with care. In Figure~\ref{fig:6} we plot the correlation function evolution for all our reionization histories (as in Figure~\ref{fig:3} for the cross-power spectra). These results are in qualitative agreement with previous work, such as \citet{2018MNRAS.479L.129H,2017ApJ...836..176H,2018MNRAS.479.2754K}\footnote{Although we note that when making comparisons, one must be careful of units. In particular we correlate the 21-cm brightness temperature itself (and not the contrast from the mean), and hence retain units of mK in the cross-correlation function.}.

The correlation functions can be thought of as the average 21-cm brightness profiles around LAEs. At the intermediate redshifts we find that there is an anticorrelation signal for scales $r<10$ cMpc/h, which dies off at larger scales. This is driven by the presence of ionized bubbles around LAEs, which sit in the overdensities that power reionization. We note that it might be possible to use the scale at which the correlation function dies to zero to estimate the typical ionized bubble size; further exploration of this is left to future work.

  Conversely at very small scales, we see the hint of an upturn in the correlation signal due to residual/self-shielded neutral hydrogen in the overdensities that host the LAEs. In the $z=5.756$ panels we see that the original models show no correlation signal, whilst the delayed-end models show a slight anti-correlation at intermediate scales. Reionization has not fully completed in these models, and hence the ionized bubbles have not fully percolated.

  \begin{figure*}
  \includegraphics[width=\textwidth]{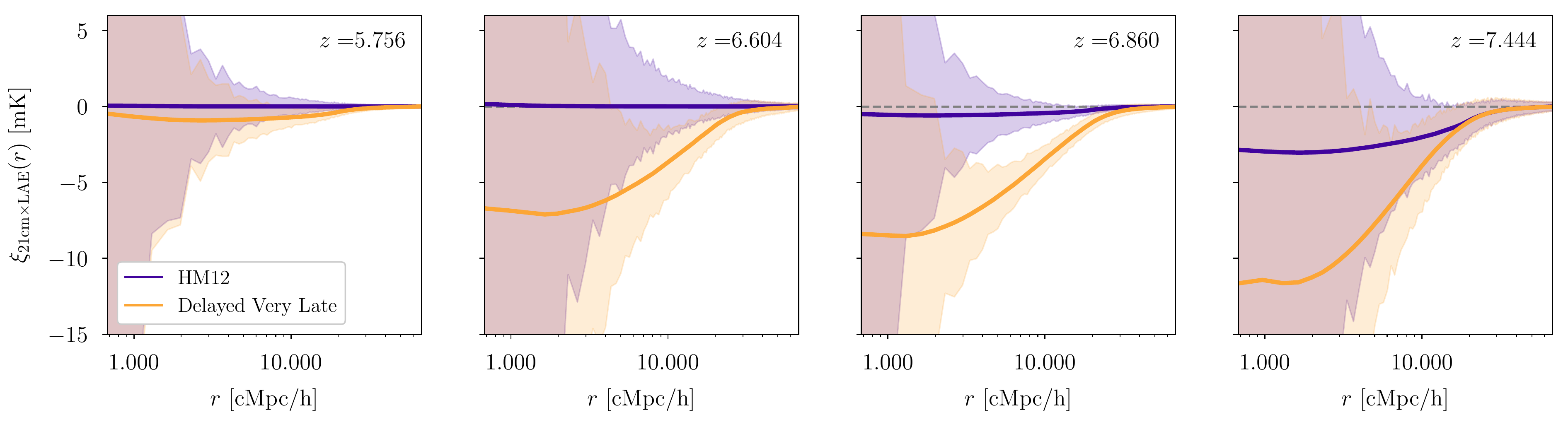}
  \caption{As in Figure~\ref{fig:5}, distinguishing reionization histories using a PFS-SKA survey: a comparison of the HM12 (purple) and delayed-end Very Late (orange) reionization histories. Here we plot the cross-correlation function, with shading indicating the 1-$\sigma$ error predicted for the PFS-SKA survey.}
  \label{fig:7}
\end{figure*}

In Figure~\ref{fig:7} we plot a comparison of the correlation functions for the HM12 and delayed-end Very Late models, showing the predicted 1-$\sigma$ errors for a PFS-SKA survey as in Figure~\ref{fig:5}. The errors have been propagated numerically using a Monte Carlo sampling of the power spectrum, assuming Gaussian statistics. We note that the chosen observational limits (see Table~\ref{tab:thresholds}) result in a larger number of LAEs at $z=6.9$ than at the lower redshifts, which reduces the shot noise at that redshift, and hence the 1-$\sigma$ errors are narrower in the $z=6.9$ panel.

\section{The effect of self-shielding}
  \label{appendix:d}
\begin{figure*}
\includegraphics[width=\textwidth]{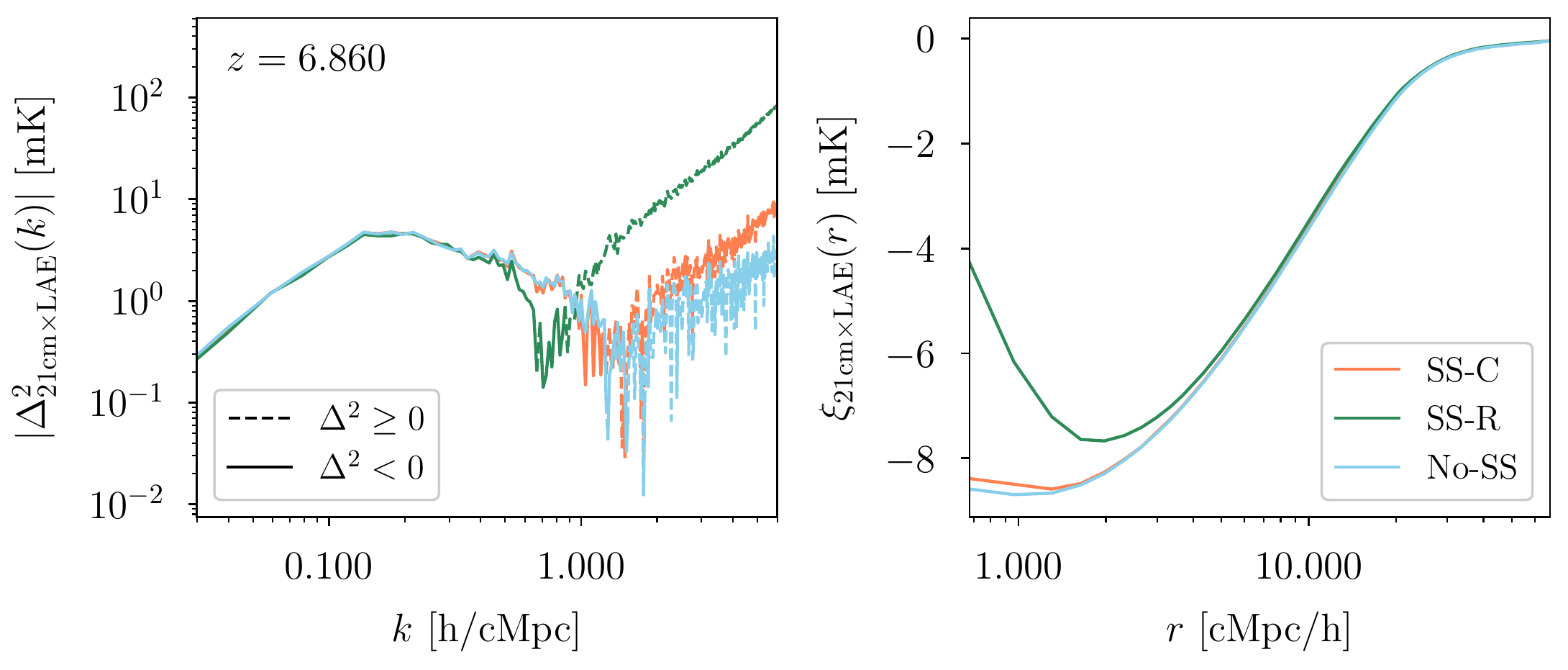}
\caption{The effect of self-shielding on the cross-power spectrum (left, as in Figure~\ref{fig:3}) and cross-correlation function (right, as in Figure~\ref{fig:6}) at $z=6.860$ for the Very Late reionization history. We show our fidicual self-shielding prescription derived from \citet{2017arXiv170706993C} in red, the \citet{2013MNRAS.430.2427R} prescription in green, and using no self-shielding prescription in blue.}
\label{fig:ss}
\end{figure*}

In this section we explore the dependence of our results on the assumed self-shielding prescription. The small-scale correlations between LAEs and the 21-cm brightness depends on the amount of neutral gas in the vicinities of host haloes. The extent of the self-shielding in gas close to the LAE can effect both how bright the gas is in 21-cm emission, but also how much the Ly$\alpha$ emission from the galaxies is attenuated. Hence if there is more self-shielding we can expect to see an increase in the correlation signal\footnote{As we found in \citet{2018MNRAS.tmp.1485W,2019MNRAS.485.1350W}, the brightest LAEs are found in the most massive haloes, where we also find significant amounts of self-shielded neutral gas. If we increase the amount of self-shielding then only the brightest LAEs -- whose positions are highly correlated with the bright self-shielded gas -- remain visible, thus enhancing the clustering.}.

In Figure~\ref{fig:ss} we test the effect of varying our self-shielding prescription on the cross-power spectrum and cross-correlation function. We note that the self-shielding prescription affects both the 21-cm brightness calculation and the LAE transmission, so there is a coupled changed in both of our observables. In this plot we show our fiducial self-shielding prescription derived from \citet{2017arXiv170706993C} in red, the \citet{2013MNRAS.430.2427R} prescription in green, and using no prescription in blue. Considering the power spectrum we see that on large scales (below $k < 0.3$ h/cMpc), varying the self-shielding assumptions does not have any effect. As we found in section~\ref{sec:results}, current and upcoming surveys are most sensitive to the reionization signal on these large scales, hence our forecasts for measuring the 21-cm LAE cross-power spectrum are robust to the self-shielding modelling choices. However on smaller scales (above $k \geq 0.3$ h/cMpc) we see that the self-shielding modelling has a large effect on the resulting power spectrum. In particular, although across all prescriptions the shape of the power spectrum remains a power law with similar index, the amplitude is a strong function of the self-shielding. As a result, the turnover scale is also dependent on the self-shielding modelling. In the right panel of Figure~\ref{fig:ss} we show the cross-correlation function. Here we see that the amount of self-shielding controls the extent to which the correlation function turns over at small scales.


\bsp	
\label{lastpage}
\end{document}